\documentclass[aps,prb,twocolumn]{revtex4}
\usepackage{amsfonts}
\usepackage{amscd}
\usepackage{bm}
\usepackage{amstext}
\usepackage{amsmath}
\usepackage{amssymb}
\usepackage{mathptmx}
\usepackage{enumerate}
\usepackage{graphicx}
\usepackage{color}
\usepackage{ifpdf}
\usepackage{epstopdf}
\DeclareGraphicsExtensions{.eps, .pdf, .jpg, .tif}
\usepackage[dvips]{hyperref}
\usepackage{float}
\usepackage{bookmath}
\usepackage{mymatrices}
\usepackage{feynmp}
\def\mF{m_{\text{F}}}
\def\mH{M_{\text{H}}}
\def\mNG{M_{\text{NG}}}
\def\mAH#1#2{M_{\text{$#1,#2$}}}
\def\ns{\negthinspace}

\def\nt{\negthickspace}
\def\intt{\negthickspace\int\negthickspace}

\def\t#1/#2{{\rm t}_{#1}^{#2}}

\def\e0{\ve^{(0)}}
\def\eplus{\ve^{(+)}}
\def\eminus{\ve^{(-)}}
\def\G{{\mathsf G}}
\def\E{\omega}
\def\betaB{\beta_{\text{B}}}
\def\betaBW{\beta_{12}+\nicefrac{1}{3}\beta_{345}}
\def\charge{\mbox{\small c}}
\begin{document}
\title{On Nambu's Fermion-Boson Relations for Superfluid \He}
\author{J. A. Sauls} 
\affiliation{Department of Physics and Astronomy, 
             Northwestern University, Evanston, IL 60208 USA}
\email{sauls@northwestern.edu}
\author{Takeshi Mizushima} 
\affiliation{Department of Materials Engineering Science, Osaka University, 
		Toyonaka, Osaka 560-8531, Japan}
\date{February 14, 2017}
%---------------------------------------------------------------------------------------------------
\begin{abstract}

Superfluid \He\ is a spin-triplet ($S=1$), p-wave ($L=1$) BCS condensate of Cooper pairs with total
angular momentum $J=0$ in the ground state. In addition to the breaking of $\point{U(1)}{\ns}$ gauge
symmetry, separate spin or orbital rotation symmetry is broken to the maximal sub-group,
$\spin\times\orbital\rightarrow\point{SO(3)}{\text{J}}$.
The Fermions acquire mass, $\mF\equiv\Delta$, where $\Delta$ is the BCS gap.
There are also 18 Bosonic excitations - 4 Nambu-Goldstone (NG) modes and 14 massive amplitude Higgs
(AH) modes. The Bosonic modes are labeled by the total angular momentum, $J\in\{0,1,2\}$, and parity
under particle-hole symmetry, $c=\pm 1$.
For each pair of angular momentum quantum numbers, $J,J_z$, there are two Bosonic partners with
$c=\pm 1$.
Based this spectrum Nambu proposed a sum rule connecting the Fermion and Boson masses for BCS type
theories, which for \Heb\ is $\mAH{J}{+}^2 + \mAH{J}{-}^2 = 4\mF^2$ for each family of Bosonic modes
labeled by $J$, where $\mAH{J}{c}$ is the mass of the Bosonic mode with quantum numbers
$(J,c)$.
Nambu's sum rule (NSR) has recently been discussed in the context of Nambu-Jona-Lasinio models for physics
beyond the standard model to speculate on possible partners to the recently discovered Higgs Boson
at higher energies.
Here we point out that Nambu's Fermion-Boson mass relations are not exact. Corrections to the
Bosonic masses from (i) leading order strong-coupling corrections to BCS theory, and (ii) 
polarization of the parent Fermionic vacuum lead to violations of the sum-rule.
Results for these mass corrections are given in both the $T\rightarrow 0$ and $T\rightarrow T_c$ limits. 
We also discuss experimental results, and theoretical analysis, for the masses of the $J^{\charge}=2^{\pm}$ 
Higgs modes and the magnitude of the violation of the NSR.
\end{abstract}
\maketitle
%----------------------------------------------------------------------------------------------
\section{Introduction}
\vspace*{-3mm}

One of the key features of spontaneous symmetry breaking in condensed matter and quantum
field theory is the emergence of new elementary quanta - phonons in crystalline solids,
magnons in ferromagnets, the Higgs and gauge bosons of the standard model.
In the latter example, spontaneous symmetry breaking (SSB) in the BCS theory of superconductivity
played an important role in theoretical models for the mass spectrum of elementary
particles.\cite{nam61a,and63,hig64}
In BCS superfluids the binding of Fermions into Cooper pairs leads to an energy gap, $\Delta$,
in the Fermion spectrum, i.e. Fermions in the broken symmetry phase (Bogoliubov quasiparticles)
acquire a mass $\mF=\Delta$, while condensation of Cooper pairs leads to the breaking of global
$\point{U(1)}{\ns}$ gauge symmetry, the generator being particle number.
The latter also implies that the Bogoliubov Fermions are no longer particle number (Fermion ``charge'')
eigenstates, but coherent superpositions of normal-state particles and holes. Charge conservation is
ensured by an additional contribution to the charge current - a collective mode of the broken
symmetry phase. This massless Bosonic excitation of the phase of condensate
amplitude\cite{and58,bog58a} is the Nambu-Goldstone (NG) mode associated with broken
$\point{U(1)}{\ns}$ symmetry, and is manifest as a phonon in neutral superfluid \He. 

\vspace*{-3mm}
\section{Nambu's Mass Relations}\medskip
\vspace*{-3mm}

Nambu and Jona-Lasinio's development of a dynamical theory for the masses of elementary particles
\cite{nam61a} was influenced by the BCS theory of superconductivity, and particularly Bogoliubov
\cite{bog58a}, Valatin \cite{val58} and Anderson's \cite{and58a,and63} contributions on the
excitation spectrum of Fermions and the collective excitations (Bosonic) associated with broken
gauge symmetry.\cite{nam09} BCS-type theories, including the NJL theory, imply a connection between
the masses of the emergent Fermionic and Bosonic excitations.
In the case of conventional BCS theory there are two Bosonic modes - the phase mode and the
amplitude mode with mass $\mH = 2\Delta$. The phase mode, discussed independently by Anderson and
Bogoliubov, is the massless NG mode ($M_{\text{NG}}=0$),
while the amplitude mode is the Higgs Boson of BCS theory.\cite{hig64,lit81} 
This doubling of the Bosonic spectrum reflects a discrete symmetry under charge conjugation, $C$, 
(i.e. ``particle $\leftrightarrow$ hole'' symmetry) of the symmetry un-broken Fermionic 
vacuum,\cite{ser83a,sau91} and is characteristic of spontaneous symmetry breaking of the BCS type, 
including BCS systems with more complex symmetry breaking phase transitions. 
In particular, the amplitude (phase) mode has even (odd)
parity with respect to charge conjugation.\cite{ser83a}
Furthermore, the masses of the Fermions and the Bosons obey 
the sum rule $M_{\text{NG}}^2 + M_{\text{H}}^2 = (2m_{\text{F}})^2$.

Nambu argued that similar sum rules apply to a broad class of BCS type theories - from nuclear 
structure and QCD to exotic pairing in condensed matter systems - that exhibit complex symmetry 
breaking.\cite{nam85} 
The ground state of superfluid \He\ provides the paradigm. Superfluid \Heb\ is a
condensate of p-wave ($L=1$), spin-triplet ($S=1$) Cooper pairs with total angular momentum $J=0$.
Thus, in addition to the breaking of $\point{U(1)}{\ns}$, the symmetry of the normal quantum liquid
with respect to separate spin or orbital rotations is broken to the maximal sub-group,
$\spin\times\orbital\rightarrow\point{SO(3)}{\text{J}}$. 
The Fermion spectrum is isotropic and gapped with mass determined by the binding energy of Cooper
pairs, $\mF=\Delta$. However, there are now 18 Bosonic excitations - 4 NG modes and 14 massive Higgs modes.
The Bosonic modes are organized into six multiplets labeled by $J^{\charge}$ - total angular
momentum, $J\in\{0,1,2\}$, and parity under charge conjugation (particle $\leftrightarrow$ hole),
$\charge=\pm 1$.\footnote{Modes with $J\ge 3$ are also possible if we include sub-dominant pairing
interactions with orbital angular momenta $\ell\ge 3$, even if the ground state is $\ell=1$,
$S=1$ and $J=0$.\cite{sau81}}
For each $J$ there are $2J+1$ degenerate states with angular momentum projection $m=-J,\ldots,+J$,
and for each pair of values of $J,m$ there are two Bosonic modes with $\charge=\pm 1$.

The $J=0$ modes are the NG mode associated with broken $\point{U(1)}{\negthickspace}$ symmetry
($J^{\charge}=0^{-}$) and the Higgs mode ($J^{\charge}=0^{+}$), which has the same quantum numbers as the B-phase
vacuum, i.e. the condensate of ground state Cooper pairs.
There are six $J=1$ modes: 3 NG modes ($J^{\charge}=1^+$) corresponding to the degeneracy of the B-phase
ground state with respect to \emph{relative} spin-orbit rotations, and 3 Higgs modes ($J^{\charge}=1^{-}$)
with masses $M_{1,-}=2\Delta$.\cite{mck93}
Finally, there are ten modes with $J=2$, all of which are Higgs modes with masses $M_{2,\pm} <
2\Delta$, with original calculations giving $M_{2,+}=\sqrt\frac{2}{5}\,\,2\Delta$ and
$M_{2,-}=\sqrt\frac{3}{5}\,\,2\Delta$.\cite{vdo63,mak74,wol76,bru80}
Nambu noted that all three multiplets obey a sum rule connecting the masses of the conjugate Bosonic
modes and the Fermionic mass,\cite{nam85}
\be
M_{J,+}^2 + M_{J,-}^2 = (2m_{\text{F}})^2\,,\quad J \in \{0,1,2\}
\,,
\ee
and suggested that such Fermion-Boson relations are generic to BCS-type NJL models in which both
Fermion and Boson excitations originate from interactions between massless progenitor Fermions and
spontaneous symmetry breaking (see also Ref. \onlinecite{vol13}). Nambu further speculated that 
these Fermion-Boson mass relations reflected a hidden supersymmetry in class of BCS-NJL models,\cite{nam85} 
and in the case of of conventional s-wave, spin-singlet BCS superconductivity was able to construct
a supersymmetric representation for the static part of the effective Hamiltonian, $H_{\mbox{s}}$, and 
identify the superalgebra as $\point{su(2/1)}{\ns}$. 
The Fermion operators in Nambu's construction factorize $H_{\mbox{s}}$, and provide 
ladder operators connecting the Fermionic and Bosonic sectors of the spectrum, and generate the 
Fermion-Boson mass relations: $\mNG=0$, $\mF=\Delta$, and $\mH=2\Delta$.\cite{nam95}$^{,}$\footnote{A 
similar analysis for \Heb\ should be possible, but the construction of the ladder operators 
and the identification of the superalgebra for a supersymmetric representation of the Hamiltonian 
for the B-phase of \He\ is a future challenge.}

Recently, Volovik and Zubkov argued that the Nambu sum rule (NSR) for \Heb\ follows from the
\emph{symmetry} of the B-phase vacuum and the quantum numbers $(J,m)$ (\emph{c.f.} Sec. 2.2 of 
Ref. \onlinecite{vol14}).
Based on the NSR for a NJL-type theory of top quark condensation, the authors suggest the possibility that 
there may be a partner to the Higgs Boson with a mass of $125\,\mbox{GeV}$ - e.g. a Higgs partner near 
$270\,\mbox{GeV}$,\cite{vol13,vol14} analogous to the Higgs partners for the $J=2$ Bosonic spectrum of \Heb.
Here we point out that estimates of the mass of a Higgs partner based on such sum rules may be
imprecise because the NSR is generally violated. The origins of the violation of the NSR contain
detailed information about the parent Fermionic vacuum. 
While one might expect that the masses for the $J$ multiplets to be protected by the residual symmetry
of the broken-symmetry vacuum state, it is generally not the case.
As a result the NSR is not exact, particularly for BCS-type theories with multiplets of NG and 
Higgs Bosons with quantum numbers that are distinct from that of the broken symmetry vacuum state.
We discuss the violations of the NSR for the case of \Heb\ in two limits: (i) time-dependent 
Ginzburg-Landau (TDGL) theory appropriate for $T\lesssim T_c$ and (ii) a dynamical theory for 
coupled Fermionic and Bosonic excitations of \Heb\ within the BCS theory for p-wave, spin-triplet 
pairing (i.e. one-loop approximation to the self-energy) for temperature $T\rightarrow 0$.
In particular, interactions between the progenitor Fermions, combined with vacuum polarization, 
lead to mass shifts of the Higgs modes whose quantum numbers differ from the broken symmetry vacuum 
state, e.g. the $J^{\charge}=1^{\pm}$ and $J^{\charge}=2^{\pm}$ modes of \Heb, and thus to 
violations of the Nambu sum rule.
Explicit results for these mass corrections are derived in both the $T\rightarrow 0$ and
$T\rightarrow T_c$ limits.

In Secs. \ref{sec-GL} and \ref{sec-TDGL} we introduce a Lagrangian for the Bosonic modes of a spin-triplet, 
p-wave BCS condensate based on a time-dependent extension of Ginzburg-Landau theory (TDGL).
This allows us to identify and calculate the Bosonic spectrum for \He, and to quantify strong-coupling 
corrections to the Bosonic masses in the limit $T\rightarrow T_c$.
In particular, strong-coupling feedback 
(i.e. next-to-leading order loop corrections) leads to mass shifts, and thus violations of the NSR.
We also obtain a formula for the mass of the $J^{\charge}=2^{-}$ mode in the GL limit that could
provide a direct determination of the GL strong-coupling parameter $\beta_1$ from measurements of
the $J^{\charge}=2^{-}$ mode via ultrasound spectroscopy.

At low temperatures strong-coupling feedback corrections are suppressed.
However in Sec. \ref{sec-Polarization} we show that vacuum polarization and four-Fermion interactions, 
in both the particle-hole (Landau) and the particle-particle (Cooper) channels, lead to substantial mass
corrections for $T\ll T_c$, and in some cases strong violations of the NSR.
We discuss experimental measurements for the masses of the $J^{\charge}=2^{\pm}$ modes, and compare
the observed mass shifts with theoretical calculations of the polarization corrections to the masses
from interactions in the Landau and Cooper channels.

\vspace*{-3mm}
\section{Ginzburg-Landau Functional}\label{sec-GL}
\vspace*{-3mm}

We start from a Ginzburg-Landau (GL) functional applicable to p-wave, spin-triplet pairing beyond
the weak-coupling BCS limit, and use this formulation to obtain an effective Lagrangian for the
Bosonic fluctuations of superfluid \Heb\ in the strong-coupling limit.
The GL theory for superfluid \He\ was developed by several authors.\cite{mer73,bri73,leg75} We
follow the notation Ref. \onlinecite{rai76} which provides the bridge between the GL theory and the
microscopic theory of leading order strong-coupling effects.
The order parameter is identified with the mean-field pairing self-energy, $\hDelta(\vp)$, which is
a $2\times 2$ matrix of the spin components of the pairing amplitude.
For p-wave, spin-triplet condensates the order parameter is symmetric in spin-space, $\hDelta(\vp) =
(i\sigma_{\alpha}\sigma_y)\,A_{\alpha i}\,(\hvp)_i$, and parametrized by a $3\times 3$ complex
matrix, $A_{\alpha i}$, that transforms as a vector with respect to index $\alpha=\{x',y',z'\}$
under spin rotations, and, separately, as a vector with respect to index $i=\{x,y,z\}$ under orbital
rotations. This representation for the order parameter provides us with a basis for an irreducible
representation of the maximal symmetry group of normal \He,
$\G=\spin\times\orbital\times\gauge\times\parity\times\time$.
The GL free energy functional is then constructed from products of $A_{\alpha i}$ and its
derivatives, $\partial_j A_{\alpha i}$, that are invariant under $\G$. 
The general form for the GL functional for the condensation energy and gradient energy is
\ber
\cF[A]
=
\int_{V}dV\,\{U(A) + W(\partial A)\}
\,,
\eer
where
\ber
\label{eq-GL_P-wave}
U \nt\nt &=& \nt\nt
         \alpha(T)\,Tr\left(A A^{\dagger}\right) 
\nt
        +\beta_{1} \left|Tr(A A^{T})\right|^{2}
\nt
		+\beta_{2} \left[Tr(A A^{\dagger})\right]^{2}  
\\
\nt\nt
&+&
\nt\nt
		 \beta_{3} Tr\left[A A^{T} (A A^{T})^{*}\right] 
\nt\ns
		+\beta_{4} Tr\left[(A A^{\dagger})^{2}\right]
\nt\ns
		+\beta_{5} Tr\left[A A^{\dagger} (A A^{\dagger})^{*}\right] 
\,.
\nonumber
\eer
are the six invariants for the condensation energy density, and
\be
W
\nt
=
\nt
		 K_{1}
\partial_{j}A_{\alpha i} \partial_{j}A_{\alpha i}^{*}
\nt
		+K_{2}
\partial_{i}A_{\alpha i} \partial_{j}A_{\alpha j}^{*}
\nt
		+K_{3}
\partial_{j}A_{\alpha i} \partial_{i}A_{\alpha j}^{*}
\,,
\label{eq-GL_gradients}
\ee
are the three second-order invariants for the gradient energy.

Weak-coupling BCS theory can be formulated at all temperatures in terms of a stationary functional 
of $\hat{\Delta}(\vp)$,\cite{ser83,ali11}
which depends on material parameters of the parent Fermionic ground state:
$N(0)=k_{f}^{3}/2\pi^2\,v_{f} p_{f}$ is the single-spin quasiparticle density of states at the
Fermi surface, expressed in terms of the Fermi velocity, $v_f$, Fermi momentum and Fermi wave number,
$p_f=\hbar k_f$.
The GL limit of the weak-coupling functional can be expressed in the form of Eqs. \ref{eq-GL_P-wave} 
and \ref{eq-GL_gradients} with the following material parameters,
\ber
\alpha(T) &=& \frac{1}{3}N(0)(T/T_{c}-1)
\,,\quad
\beta^{\text{wc}}_{1} \equiv \frac{7\zeta(3)}{240} \frac{N(0)}{(\pi k_{\text{B}}T_{c})^2}
\,,
\label{eq-alpha_wc}
\\
2\beta^{\text{wc}}_{1} 
&=& -\beta^{\text{wc}}_{2} 
 =  -\beta^{\text{wc}}_{3} 
 =  -\beta^{\text{wc}}_{4} 
 =  +\beta^{\text{wc}}_{5}
 =  -2\beta_{\text{wc}}
\,.
\label{eq-betas_wc}
\eer

\begin{figure}[t]
\begin{center}
\includegraphics[width=\linewidth]{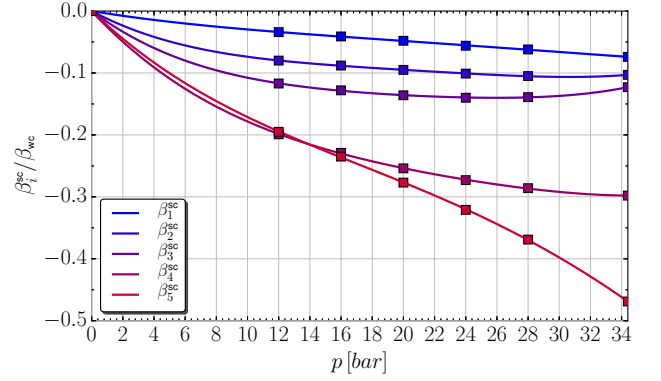}
\caption{Strong-coupling corrections ($\beta^{\text{sc}}_i\equiv\beta_i-\beta^{\text{wc}}_i$) to 
                 the GL $\beta$ parameters interpolated from the 
		 results of Ref. \onlinecite{sau81b} [data squares]. The $\beta^{\text{sc}}_{i}$ are 
		 extrapolated below $P=12$ bar to weak-coupling ($\beta^{\text{sc}}_{i}=0$) 
		 at $p=0$ bar.}
\label{fig-betas_SS}
\end{center}
\end{figure}

Strong-coupling corrections to the weak-coupling GL $\beta$ parameters based on the leading-order
expansion of Rainer and Serene\cite{rai76} were calculated and reported in Ref. \onlinecite{sau81b}
for quasiparticle scattering that is dominated by ferromagnetic spin fluctuation exchange. The
results for the strong-coupling corrections to the weak-coupling $\beta^{\text{wc}}_{i}$ are
extrapolated to all pressures as shown in Fig. \ref{fig-betas_SS}, with $p=0\,\mbox{bar}$
corresponding to weak-coupling.

The weak-coupling form of the gradient energy in Eq. \ref{eq-GL_gradients} is similarly obtained
with the gradient coefficients given by
\ber
K^{\text{wc}}_{1} 
&=& 
K^{\text{wc}}_{2} = K^{\text{wc}}_{3} = \frac{1}{5} N(0)\,\xi_{\text{GL}}^2
\,,
\label{eq-Kappas_wc}
\\
\xi_{\text{GL}} 
&=& 
\sqrt{\frac{7\zeta(3)}{12}}\frac{\hbar v_{f}}{2\pi k_{B} T_{c}}
\,.
\label{eq-magnitudes_wc}
\eer
The Cooper pair correlation length, $\xi_{\text{GL}}$, varies from $\xi_{\text{GL}}\simeq 650$ \AA{} 
at $p = 0\,\mbox{bar}$ to $\xi_{\text{GL}}\simeq 150$ \AA{} at $p = 34\,\mbox{bar}$.

The Balian-Werthamer (BW) state defined by
\vspace*{-3mm}
\be
A^{\text{BW}}_{\alpha i} = \frac{\Delta}{\sqrt{3}}\,e^{i\varphi}R[\vec{\vartheta}]_{\alpha i}
\,,
\ee
where $R[\vec\vartheta]$ is an orthogonal matrix,
minimizes the GL functional for $\Delta^2 = -\alpha(T)/2\beta_{\text{B}}$, with 
$\beta_{\text{B}}=\beta_{12}+\frac{1}{3}\beta_{345}$, in the weak-coupling limit, 
$\beta^{\text{wc}}_{\text{B}}=\nicefrac{5}{6}\beta^{\text{wc}}$, and 
for all pressures $P<P_{\text{PCP}}\approx 21\,\mbox{bar}$.
Note that the amplitude of the order parameter, $\Delta$, is fixed at the minimum of the effective
potential. However, the phase, $\varphi$, and the orthogonal matrix, $R[\vec\vartheta]$, parametrized by
a rotation angle $\vartheta$ about an axis of rotation, $\hat\vn$, define the degeneracy space of
the B-phase.
In particular, 
\vspace*{-3mm}
\be
B_{\alpha i}\equiv\frac{\Delta}{\sqrt{3}}\,\delta_{\alpha i}
\,,
\ee
corresponding to pairs with $L=1$, $S=1$ and $J=0$ is degenerate with states obtained by any
relative rotation, $R[\vec\vartheta]$, of the spin and orbital coordinates combined with a gauge
transformation, $e^{i\varphi}$.
Since the GL functional defined by Eqs. \ref{eq-GL_P-wave} and \ref{eq-GL_gradients} is invariant under 
these operations we can use the $J=0$ BW state as the reference ground state.

%------------------------------------- Strong-coupling Betas-SS  --------------------------
%----------------------------------------------------------------------------------------------

\vspace*{-3mm}
\section{Time Dependent GL Theory}\label{sec-TDGL}
\vspace*{-3mm}

Bosonic excitations of the BW ground state are represented by space-time fluctuations of the pairing
amplitude: $\cD_{\alpha i}(\vr,t) = A_{\alpha i}(\vr,t) - B_{\alpha i}$.
The potential energy for these fluctuations is obtained by expanding the GL
functional to 2$^{nd}$ order in the fluctuations $\cD(\vr,t)$: $\cU[\cD] =
\cF[A]-\cF[B]$.\cite{vol84a,the88}
Time-dependent fluctuations, $\dot{\cD_{\alpha i}}=\partial_t\cD_{\alpha i}$, lead to an
additional invariant, $\cK = \tau\int_{\text{V}}dV\,\dot\cD_{\alpha i}\dot\cD_{\alpha i}^{*}$,
where $\tau$ is the effective inertia for Cooper pair fluctuations.\footnote{We have omitted
the invariant that is first-order in time derivatives, $\Tr{\dot\cD\cD^{\dagger}} - h.c.$.
This invariant is odd under charge conjugation, and thus has a small, but non-zero prefactor, 
only because of the weak violation of particle-hole symmetry of the normal Fermionic vacuum.}
The Lagrangian for the Bosonic excitations, $\cL=\cK-\cU$, takes the form,
\ber
\label{eq-Lagrangian}
\cL
\nt = \nt
\intt\ns dV\nt
&\Bigg\{&\nt
\tau\,\Tr{\dot\cD\dot\cD^{\dag}}
\ns - \ns
\alpha\,\Tr{\cD\cD^{\dag}}
\ns - \ns
\sum_{p=1}^{5}\beta_{p\,}u_{p}(\cD)
\nonumber\\
&\ns - \ns&
\sum_{l=1}^{3}\,K_{l}\,v_{l}(\partial\cD)
\ns - \ns
\left(\eta_{\alpha i}\cD_{\alpha i}^{*} + \eta_{\alpha i}^{*}\cD_{\alpha i}\right)
\nt\Bigg\} 
\,.
\eer
The terms $u_{p}(\cD)$ are the effective potentials corresponding to fluctuations, $\cD$, 
relative to the BW ground state, which to quadratic order in $\cD$ are given in Eqs.
\ref{eq-TDGL_potential-1}-\ref{eq-TDGL_potential-5} of the Appendix.
The terms, $w_{l}(\partial\cD)$, are obtained from Eq. \ref{eq-GL_gradients} with
$A\rightarrow\cD$, and the last pair of terms in Eq. \ref{eq-Lagrangian} represent
an effective external source potential for Cooper pair fluctuations.

The Euler-Lagrange equations,
\be
\fder{\cL}{\cD_{\alpha i}^{*}} 
- 
\pder{}{t}\fder{\cL}{\dot\cD_{\alpha i}^{*}}
- 
\pder{}{x_j}\fder{\cL}{\partial_{j}\cD_{\alpha i}^{*}}
=0
\,,
\ee
reduce to field equations for the Cooper-pair fluctuations,
\be\label{eq-TDGL_equations}
\hspace*{-3mm}
\tau\ddot\cD_{\alpha i} \nt-\nt \vert\alpha\vert\cD_{\alpha i}
\ns+\ns
\Delta^2\ns\sum_{a=1}^{5}\ns\beta_a\,\pder{\bar{u}_a}{\cD_{\alpha i}^{*}}
\ns-\ns
\nt\sum_{a=1}^{3}\nt K_a\,\partial_k\pder{\bar{v}_a}{[\partial_k\cD_{\alpha i}^{*}]}
\nt=\nt \eta_{\alpha i}
\,.
\ee

The field equations reduce to coupled equations for pair fluctuation modes of wavelength $\vq$:
$\cD_{\alpha i}(\vr,t)\rightarrow \cD_{\alpha i}(\vq;t)e^{i\vq\cdot\vr}$.
Furthermore, the BW ground state is invariant under joint spin- and orbital rotations. Thus,
the $\vq=0$ Bosonic excitations can be labeled by the quantum numbers, $J$, and
$m\in\{-J,\ldots,+J\}$ for the total angular momentum and its projection along a fixed
quantization axis, $\hvz$.
The dynamical equations for the Bosonic modes decouple when expressed in terms of
spherical tensors that form bases for representations of $\point{SO(3)}{J}$
for $J=0,1,2$,
\be
\cD_{\alpha i}(\vr,t) = \sum_{J,m}\,D_{J,m}(\vr,t)\,t^{(J,m)}_{\alpha i}
\,,
\ee
where the set of nine spherical tensors defined in Table \ref{table-tensors} 
(i) span the space
of rank-two tensors, (ii) form irreducible representations of $\point{SO(3)}{J}$ and (iii)
satisfy the orthonormality conditions,
\be
\Tr{\widehat{t}^{(J,m)^\dag}\,\widehat{t}^{(J',m')}} = \delta_{J,J'}\,\delta_{m,m'}
\,.
\ee
%

%~~~~~~~~~~~~~~~~~~~~~~~~~~~~~~~~~~~~~~~~~~~~~~~~~~~~~~~~~~~~~~~~~~~~~~~~~~~~~~~~~
% Table of Spherical Tensors and Spherical Harmonics for J=0,1,2

\begin{center}
\begin{table}[t]
\begin{tabular}{l|r|c|l} %|l}
$J$&$M$ 
	& $\quad\qquad t^{(J,M)}_{ij}$
	& $\quad \cY_{Jm}(\hat\vp)$ 
\\ 
\hline
\hline
$0$ & $0$	
	& $\frac{1}{\sqrt{3}} \delta_{ij}$
	& $1$
\\ 
\hline
%	J=1
& $+1$	
	& $\sqrt{3}\,\epsilon_{ijk}\eplus_k$
	& $-\sqrt{\frac{3}{2}}\,\hat\vp_+$
\\ 
$1$ & $0$	
	& $\sqrt{3}\,\epsilon_{ijk}\e0_k$
	& $+\sqrt{3}\,\hat\vp_z $
\\ 
& $-1$	
	& $\sqrt{3}\,\epsilon_{ijk}\eminus_k$
	& $+\sqrt{\frac{3}{2}}\,\hat\vp_-$
\\ 
\hline
%	J=2 
& $+2$	
	& $\;\eplus_i\eplus_j$
	& $+\sqrt{\frac{15}{8}}\,\hat\vp_+^2$
\\ 
& $+1$	
	& $\sqrt{\frac{1}{2}}\left(\e0_i\eplus_j+\eplus_i\e0_j\right)$
	& $-\sqrt{\frac{15}{2}}\,\hat\vp_z\hat\vp_+$
\\ 
$2$ & $0$	
	& $\sqrt{\frac{3}{2}}\left(\e0_i\e0_j - \frac{1}{3}\delta_{ij}\right)$
	& $+\sqrt{\frac{5}{4}}\left(3\hat\vp_z^2 - 1\right)$
\\ 
& $-1$	
	& $\sqrt{\frac{1}{2}}\left(\e0_i\eminus_j+\eminus_i\e0_j\right)$
	& $+\sqrt{\frac{15}{2}}\,\hat\vp_z\,\hat\vp_-$
\\ 
& $-2$	
	& $\;\eminus_i\eminus_j$
	& $+\sqrt{\frac{15}{8}}\,\hat\vp_-^2$
\\ 
\hline
\end{tabular}
\caption{Irreducible tensor representations, $\{t^{(J,M)}_{ij}\}$, of $\point{SO(3)}{J}$ for $J\le 2$, 
         and corresponding spherical harmonics, $\cY_{JM}(\hat\vp)$. The base unit vectors:
		 $\e0_i = \hat\vz_i$, 
		 $\eplus_i = -\frac{1}{\sqrt{2}}\left(\hat\vx_i + i \hat\vy_i\right)$ and
		 $\eminus_i = +\frac{1}{\sqrt{2}}\left(\hat\vx_i - i \hat\vy_i\right)$
		are orthonormal: $\ve^{(\mu)*}\cdot\ve^{(\nu)}=\delta_{\mu\nu}$.
}
\label{table-tensors}
\end{table}
\end{center}
%--------------------------------------------------------------------------------------
%
In the absence of a perturbation that breaks the rotational symmetry of the ground state, there are
$(2J+1)$ degenerate modes with spin $J$. There is, in addition, a doubling of the Bosonic modes
related to the discrete symmetry of the normal Fermionic ground state under charge conjugation.
Thus, the full set of quantum numbers for the Bosonic spectrum is $\{J,m,\charge\}$ where
$\charge=\pm 1$ is the parity of the Bosonic mode under charge conjugation. The parity eigenstates
are the linear combinations (i.e. real and imaginary amplitudes)\footnote{Note that the parity of
the modes is defined relative to that of the BW ground state, which is defined as $\charge=+1$.}
\be
D_{J,m}^{(\charge)} = (D_{J,m} + \charge\,D_{J,m}^{\dag})/2
\,.
\ee
The sources can also be expanded in this basis:
$\eta_{\alpha i} = \sum_{J,m,\charge}\eta_{J,m}^{(\charge)}\,t^{(J,m)}_{\alpha i}$.
The equations for the 18 Bosonic modes then decouple into three doublets labeled by
$J,\charge$, each of which is $2J+1$-fold degenerate as shown in Table \ref{table-modes}.

The equations of motion for the 18 Bosonic modes are obtained by projecting out the $J,m,\charge$
components of Eq. \ref{eq-TDGL_equations}. In the limit $\vq=0$ the modes decouple into three
doublets labeled by $J,\charge$, each of which is $2J+1$-fold degenerate. The dispersion of the
Bosonic modes can be calculated perturbatively to leading order in $(v_f|\vq|/\Delta)^2$. Thus, the
resulting equations of motion can be expressed as
\be
\partial_{t}^2 D_{J,m}^{(\charge)} 
+ 
\E_{J,m}^{(\charge)}(\vq)^2
\,D_{J,m}^{(\charge)}=\frac{1}{\tau}\eta_{J,m}^{(\charge)}
\,,
\ee
\vspace*{-5mm}
\be
\hspace*{-2.25cm}
\mbox{where}
\hspace*{0.5cm}
\E_{J,m}^{(\charge)}(\vq)=\sqrt{M_{J,{\charge}}^{\,2}+\left(c_{J,|m|}^{(\charge)}|\vq|\right)^2}
\ee
is the dispersion relation for Bosonic excitations with with quantum numbers $\{J,m,\charge\}$ and
$M_{J,{\charge}}$ is the corresponding excitation energy at $\vq=0$, i.e. the mass.
For $\vq\ne 0$ the degeneracy of the Bosonic spectrum is partially lifted, i.e. the velocities,
$c_{J,|m|}^{(\charge)}$, give rise to a dispersion splitting that depends on $|m|$, with
quantization axis $\vq$.\cite{vol84b,fis88a}

\vspace*{-3mm}
\subsection{$J=0$ Modes}
\vspace*{-3mm}

%------------------------------------- Table of Higss Modes  --------------------------
\begin{table}[t]
\begin{center}
\begin{tabular}{c|c|c|l}
		Mode	&	Symmetry	&	Mass	&	Name	
\\
\hline
	$D^{(+)}_{0,m}$		&	$J=0$, $\charge=+1$	&	$2\Delta$		&	Amplitude 
\\
	$D^{(-)}_{0,m}$		&	$J=0$, $\charge=-1$	&	$0$				& 	Phase Mode	
\\
\hline
	$D^{(+)}_{1,m}$		&	$J=1$, $\charge=+1$	&	$0$				&	NG Spin-Orbit Modes
\\
	$D^{(-)}_{1,m}$		&	$J=1$, $\charge=-1$	&	$2\Delta$		&	AH Spin-Orbit Modes
\\
\hline
	$D^{(+)}_{2,m}$		&	$J=2$, $\charge=+1$	&	$\sqrt{\nicefrac{8}{5}}\Delta$	
																	&	 $2^{+}$ AH Modes
\\
	$D^{(-)}_{2,m}$		&	$J=2$, $\charge=-1$	&	$\sqrt{\nicefrac{12}{5}}\Delta$	
																	&	 $2^{-}$ AH Modes
\\
\hline
\end{tabular}
\end{center}
\caption{
Bosonic Mode Spectrum for the B-Phase of \He. The masses of the modes are given for weak-coupling in
the GL limit.
}
\label{table-modes}
\end{table}
%--------------------------------------------------------------------------------------

The masses and velocities of the Bosonic modes obtained from the TDGL Lagrangian in the weak-coupling 
limit are summarized in Table \ref{table-modes}.
The $J=0$ modes correspond to the two Bosonic modes that are present for any BCS
condensate of Cooper pairs, i.e. excitations of the phase, $D_{0,0}^{(-)}$, and amplitude,
$D_{0,0}^{(+)}$, with the \emph{same internal symmetry as the condensate} of Cooper pairs.
The $J^{\charge}=0^{-}$ mode is the Anderson-Bogoliubov (AB) phase mode. 
In particular, if we consider \emph{only} fluctuations of the phase of the 
BW ground state, $A_{\alpha i}=B_{\alpha
i}\,e^{i\vartheta(\vr,t)}\approx B_{\alpha i}(1 + i\vartheta(\vr,t))$, then
$D_{0,0}^{(-)}=i\Delta\,\vartheta(\vr,t)$.
This is the massless NG mode corresponding to the broken $\point{U(1)}{}$ symmetry, with the
dispersion relation $\E_{0,0}^{(-)} = c_{0,0}|\vq|$.
Within the TDGL theory the AB mode propagates with velocity
$c_{0,0}=\sqrt{(K_{1}+\nicefrac{1}{2}K_{23})/\tau}$.
In the weak-coupling limit for the effective action derived by Bosonization of the Fermionic
action the velocity is $c_{0,0}= v_f/\sqrt{3}$,\cite{popov87}
showing that the Bosonic excitation energies are determined by the properties of the underlying
Fermionic vacuum - in this case the group velocity of normal-state Fermionic excitations at the
Fermi surface.
However, this result for the velocity of the NG phase mode is further renormalized by coupling
of the phase fluctuations to dynamical fluctuations of the underlying Fermionic vacuum which
are absent from the Bosonic action based on the TDGL Lagrangian of Eq. \ref{eq-Lagrangian}.
This coupling leads to $c_{0,0}\rightarrow c_1 + (c_0 - c_1)\,\cY(T/T_c)$, where $c_1 (c_0)$ is the
first (zero) sound velocity of the interacting \emph{normal} Fermi liquid and $\cY(T/T_c)$ measures 
the dynamical response of the condensate. In particular, $\cY\rightarrow 0$
($\cY\rightarrow 1$) for $T\rightarrow 0$ ($T\rightarrow T_c$).
This remarkable result shows that the velocity of the NG phase mode is renormalized to the
hydrodynamic sound velocity of normal \He\ at $T=0$, and that the $J=0,\charge=-1$ NG mode is
manifest in superfluid \He\ as longitudinal sound.\cite{ser74,wol77,sau00a}

The partner to the NG phase mode is the $J^{\charge}=0^{+}$ ``amplitude'' mode. This is the Higgs
Boson of superfluid \He, i.e. the Bosonic excitation of the condensate with the \emph{same internal
symmetry} ($L=1$, $S=1$, $J=0$, $\charge=+1$) as condensate of Cooper pairs that comprise the ground
state.\cite{hig64}
For this reason the renormalizations of the $J^{\charge}=0^{+}$ Bosonic mass and the mass of
Fermionic excitations of the $J^{\charge}=0^{+}$ BW state are equivalent; thus, $M_{0,+} = 2\mF$,
where $\mF=\Delta$ is the renormalized Fermionic mass in the dispersion relation for Fermionic
excitations, $E_{\vp}^2=\mF^2 + v_f^2(p-p_f)^2$.
This allows us to fix the effective inertia of the Bosonic 
fluctuations in the TDGL Lagrangian of Eq. \ref{eq-Lagrangian} for the BW ground state as 
$\tau=\betaB\equiv\beta_{12}+\nicefrac{1}{3}\beta_{345}$.
Thus, the Nambu sum rule, $M_{0,-}^{2} + M_{0,+}^{2} = 4\mF^2$, is obeyed for the $J=0$ modes.
However, strong-coupling corrections to the TDGL Lagrangian 
lead to violations of the Nambu sum rule for Bosonic excitations with $J\ne 0$. 

\vspace*{-3mm}
\subsection{Violations of the Nambu Sum Rule for $J\ne 0$}
\vspace*{-3mm}

%--------------------------  TDGL Higgs Masses vs Pressure  ------------------------------------
\begin{figure}[t]
\begin{center}
\includegraphics[width=\linewidth]{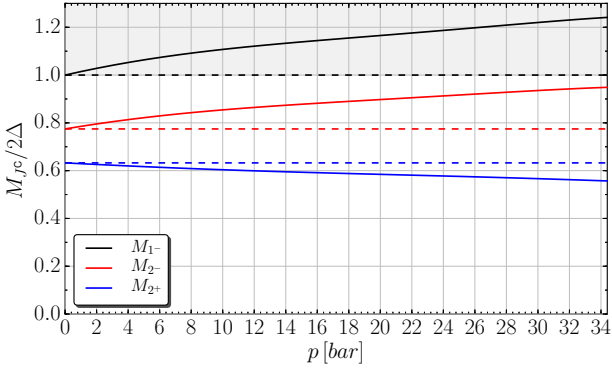}
\caption{Strong-coupling corrections to the Bosonic masses obtained from the TDGL theory
         for the GL $\beta$ parameters 
		 shown in Fig. \ref{fig-betas_SS}. The dashed lines correspond to the weak-coupling
		values for the masses.}
\label{fig-TDGL_Higgs-Masses}
\end{center}
\end{figure}
%--------------------------------------------------------------------------------------

In addition to the NG mode associated with broken $\point{U(1)}{\nt}$ symmetry, there are 3 NG modes
associated with spontaneously broken \emph{relative} spin-orbit rotation symmetry,
$\spin\times\orbital\rightarrow\point{SO(3)}{J}$. These NG modes reflect the degeneracy of the BW
ground state with respect to relative spin-orbit rotations, $\point{SO(3)}{L-S}$, whose generators
form a vector representation of $\point{SO(3)}{\ns}$. Thus, the corresponding NG modes are the
$J^{\charge}=1^{+}$ modes,
which are spin-orbit waves with excitation energies, $\E_{1,m} = c_{1,m}|\vq|$, and velocities,
$c_{1,0}=\nicefrac{1}{5}v_f$ and $c_{1,\pm 1}=\nicefrac{2}{5}v_f$ in the weak-coupling
limit.\cite{popov87} The velocities are also renormalized in the limit $T\rightarrow 0$ by the
coupling to dynamical fluctuations of the underlying Fermionic vacuum.\footnote{The weak breaking of 
\emph{relative} spin-orbit rotation symmetry by the nuclear dipolar interaction present in the
normal-state
partially lifts the degeneracy of the $J^{\charge} = 1^{+}$ NG modes, endowing the 
$m=0$ mode with a very small mass determined by the nuclear dipole energy. This is the ``Light Higgs'' 
scenario discussed by Zavjalov et al.\cite{zav16} See Sec. \ref{sec-Light_Higgs}.}

The partners to these NG modes are the $J^{\charge}=1^{-}$ Higgs modes with mass
\be
M_{1,-}=2\Delta
			   \left(\frac{-\beta_1 + \nicefrac{1}{3}(\beta_{4}-\beta_{35})}
			              {\betaBW}\right)^{\nicefrac{1}{2}}
\,,
\ee
which reduces to $M^{\text{wc}}_{1,-}=2\Delta$ in the weak-coupling limit for the GL $\beta$ parameters
(Eqs. \ref{eq-betas_wc}).
However, in the strong-coupling limit the masses of the $J^{\charge}=1^{-}$ modes deviate from
$2\mF$, which implies a violation of the NSR for the $J=1$ Bosonic modes.
Theoretical calculations of the strong-coupling $\beta$ parameters predict that
the $J^{\charge}=1^{-}$ Higgs modes 
are pushed to energies 
above the pair-breaking edge of 
$2\Delta$, as shown in Fig. \ref{fig-TDGL_Higgs-Masses}. 
This opens the possibility for the 
$J^{\charge}=1^{-}$ modes to decay into un-bound Fermion pairs. Thus, we expect the 
$J^{\charge}=1^{-}$ modes are at best resonances with finite lifetime.

For $J=2$ there are two 5-fold multiplets of Higgs modes with masses
\ber
M_{2,+}=2\Delta
	    \left(\frac{\nicefrac{1}{3}\beta_{345}}
			       {\betaBW}\right)^{\nicefrac{1}{2}}
\,,
\label{eq-TDGL_2plus-mode}
\\
M_{2,-}=2\Delta
			   \left(\frac{-\beta_1}
			              {\betaBW}\right)^{\nicefrac{1}{2}}
\,.
\label{eq-TDGL_2minus-mode}
\eer
Equation \ref{eq-TDGL_2minus-mode} provides a fifth observable that might be used to
determine GL $\beta$ parameters from independent experiments in the GL regime.\cite{cho07}
In the weak-coupling limit with $\beta_{i}$ given by Eqs. \ref{eq-betas_wc}, the
masses reduce to
$M_{2,+}^{\text{wc}}=\sqrt\frac{8}{5}\Delta$ and 
$M_{2,-}^{\text{wc}}=\sqrt\frac{12}{5}\Delta$.
Thus, the $J^{\charge}=2^{\pm}$ Higgs modes obey the NSR in the weak-coupling 
limit of the TDGL theory.\cite{nam85,vol13,vol14}

However, the NSR is violated by strong-coupling corrections to the Higgs masses, shown in Fig.
\ref{fig-TDGL_Higgs-Masses} as a function of pressure for the strong-coupling $\beta$ parameters
shown in Fig. \ref{fig-betas_SS}.
The asymmetry in the mass corrections for $M_{2,\pm}$ leads to a
sizable violation of the NSR at high pressures:
$\sum_{\charge}M_{2,\charge}^2/4\mF^2 - 1 \approx 20\,\%$ at $p=34\,\mbox{bar}$.
The violations of the NSR have the following origin: The strong-coupling Lagrangian for the Bosonic
fluctuations, Eqs. \ref{eq-Lagrangian} and
\ref{eq-TDGL_potential-1}-\ref{eq-TDGL_potential-5}, depends on the symmetry of the mode; thus, the
strong-coupling renormalization of the Higgs masses depends on $J^{\charge}$.
For the $J=0^{+}$ mode the strong-coupling renormalization of the mass is the same as that of the
$J=0^{+}$ ground state amplitude $\Delta$, and thus the Fermion mass, in which case the NSR is
satisfied even with strong-coupling corrections.
However, for modes with $J\ne 0$, the renormalization of the mass of the Higgs mode is a different
combination of the strong-coupling $\beta$'s than that which renormalizes $\Delta$, leading to
violations of the NSR.

\vspace*{-3mm}
\section{Beyond TDGL Theory}\label{sec-Polarization}
\vspace*{-3mm}

The TDGL theory is limited in its applicability because it is based on an
effective action with {\it only} Bosonic degrees of freedom. However, the parent state of a BCS condensate
is the Fermi liquid ground state (``Fermionic vacuum''). In order to calculate effects on the
Bosonic spectrum arising from ``back-action'' of the Fermionic vacuum we require a dynamical theory
that includes both Fermion and Bosonic degrees of freedom.

Microscopic formulations of the theory of collective excitations in superfluid \Heb\ were developed on the
basis of mean-field kinetic equations in Ref. \onlinecite{wol74}, Kubo theory in Refs.
\onlinecite{mak74} and \onlinecite{tew79a}, a functional integral formulation of the hydrodynamic action in Ref.
\onlinecite{bru80}, and quasi-classical transport theory in Refs. \onlinecite{sau81,moo93,mck90,mck90a}.
We highlight the coupling between Bosonic and Fermionic degrees of freedom that lead to mass
shifts of the Higgs modes. Results for the mass shifts of the $J^{\charge}=2^{\pm}$ Higgs modes 
reported in Ref. \onlinecite{sau81} are interpreted here in terms of interactions that result  
from polarization of the Fermionic vacuum by the creation of a Bosonic mode that has different
symmetry than that of the un-polarized vacuum.
The Higgs modes with different parities, $\charge=\pm 1$, also polarize the Fermionic vacuum in 
different channels, activating different interactions and leading to different mass shifts.
Thus, the violation of the $J=2$ NSR is directly related to the vacuum polarization
mass shifts for the two charge conjugation partners of the $J=2$ multiplet. 

\subsection{Particle-Hole Self Energy}

For an interacting Fermi system the two-body interaction between isolated \He\ atoms is renormalized
to effective interactions between low-energy Fermionic quasiparticles that are well defined
excitations within a low energy band near the Fermi surface, $|\varepsilon|\le\hbar\Omega_c\ll E_f$,
and thus a shell in momentum space, $\delta p \le \hbar\Omega_c/v_f$.

A disturbance of the vacuum state from that of an isotropic Fermi sea, e.g. by a
perturbation that couples to the quasiparticle states in the vicinity of the Fermi surface,
generates a polarization of the Fermionic vacuum, and a corresponding self energy correction to the
energy of a Fermionic quasiparticle.
The leading order correction is given by the combined external field, $u_{\alpha\beta}(p)$, 
plus mean-field (one loop) 
interaction energy associated with a particle-hole excitation of the Fermionic vacuum state,
%
%---------- Sigma-ph ------------------
\setlength{\unitlength}{1mm}
\begin{fmffile}{fmf_Sigma}
\input{feynman_defs.sty}
\ber
\hspace*{-5mm}
\Sigma_{\alpha\beta}(p) 
&=&\quad
%
%--------------------- External Field Energy  -------------------------------
	\parbox{22mm}{
	\begin{fmfgraph*}(18,18)
		\fmfforce{(0.10w,0.5h)}{v2}
		\fmfforce{(0.90w,0.5h)}{v4}
		\fmfforce{(0.50w,0.90h)}{v8}
		\fmfv{l=$\hspace*{0.5mm}\vspace*{-2.75mm}\noexpand\large\times$}{v8}
		\fmfforce{(0.50w,0.50h)}{v9}
		\fmfv{d.sh=circle,d.fi=0.0,d.si=0.20w,l=$\hspace*{-3mm}\vspace*{-1.0mm}\noexpand u$}{v9}
		\fmf{wiggly,l=$\noexpand u$}{v8,v9}
		\fmf{fermion}{v4,v9}
		\fmf{fermion}{v9,v2}
		\fmfv{l=\footnotesize$\noexpand p\; \alpha$,    l.a=180, l.d=0.05w}{v2}
		\fmfv{l=\footnotesize$\noexpand p\; \beta$,     l.a=0,   l.d=0.05w}{v4}
	\end{fmfgraph*}
	}
%--------------------------------------------------------------------------
\hspace*{1mm} + \hspace*{5mm}
%--------------------- Mean Field Self Energy  -------------------------------
	\parbox{22mm}{
	\begin{fmfgraph*}(18,18)
		\fmfforce{(0.10w,0.50h)}{v2}
		\fmfforce{(0.90w,0.50h)}{v4}
		\fmfforce{(0.375w,0.50h)}{v5}
		\fmfforce{(0.675w,0.50h)}{v8}
		\fmfforce{(0.375w,0.75h)}{v6}
		\fmfforce{(0.675w,0.75h)}{v7}
		\fmfforce{(0.525w,0.625h)}{v9}
		\fmfv{d.sh=square,d.fi=0.0,d.si=0.25w,
			  l=$\hspace*{-1mm}\vspace*{-3.5mm}\noexpand \Gamma^{\mbox{\tiny ph}}$}{v9}
		\fmf{fermion}{v4,v8}
		\fmf{fermion}{v5,v2}
		\fmf{double_fermion,right=3.5,width=1.25pt}{v7,v6}
		\fmfv{l=\footnotesize$\noexpand p\; \alpha$,    l.a=180, l.d=0.05w}{v2}
		\fmfv{l=\footnotesize$\noexpand p\; \beta$,     l.a=0,   l.d=0.05w}{v4}
	\end{fmfgraph*}
	}
%--------------------------------------------------------------------------
\label{eq-Sigma_one-loop}
\,.
\eer
\end{fmffile}
%---------------------------------------

\vspace*{-7mm}
\noindent The interaction between Fermionic quasiparticles shown in Eq. \ref{eq-Sigma_one-loop} is
represented by a four-point vertex that sums bare two-body interactions to all orders involving all
possible intermediate states of high-energy Fermions.
The vertex that determines the leading order quasiparticle self energy,
$\Gamma^{\text{ph}}$, defines the forward-scattering amplitude for particle and hole pairs
(Landau channel) scattering within the low-energy shell near the Fermi surface.

%---------- Gamma-ph Vertex ------------
\vspace*{-5mm}

\setlength{\unitlength}{1mm}
\begin{fmffile}{fmf_Gamma-ph}
\input{feynman_defs.sty}
\ber\label{eq_Gamma-ph}
\hspace*{-5mm}
\Gamma^{\mbox{\footnotesize ph}}_{\alpha\beta;\gamma\rho}(p,p') 
&=&\quad
%
%--------------------- Particle-Hole Vertex -------------------------------
	\parbox{22mm}{
	\begin{fmfgraph*}(18,18)
		\fmfforce{(0.01w,0.75h)}{v1}
		\fmfforce{(0.01w,0.25h)}{v2}
		\fmfforce{(0.99w,0.75h)}{v3}
		\fmfforce{(0.99w,0.25h)}{v4}
		\fmfforce{(0.375w,0.375h)}{v5}
		\fmfforce{(0.375w,0.625h)}{v6}
		\fmfforce{(0.625w,0.375h)}{v8}
		\fmfforce{(0.625w,0.625h)}{v7}
		\fmfforce{(0.50w,0.50h)}{v9}
		\fmfv{d.sh=square,d.fi=0.0,d.si=0.25w,			  
			  l=$\hspace*{-4mm}\noexpand \Gamma^{\mbox{\tiny ph}}$}{v9}
		\fmf{plain}{v5,v6}
		\fmf{plain}{v6,v7}
		\fmf{plain}{v7,v8}
		\fmf{plain}{v8,v5}
		\fmf{fermion}{v1,v6}
		\fmf{fermion}{v5,v2}
		\fmf{fermion}{v7,v3}
		\fmf{fermion}{v4,v8}
		\fmfv{l=\footnotesize$\noexpand p\; \alpha$,    l.a=180,  l.d=0.1w}{v2}
		\fmfv{l=\footnotesize$\noexpand p^{'}\gamma$,   l.a=180,  l.d=0.1w}{v1}
		\fmfv{l=\footnotesize$\noexpand p\; \beta$,     l.a=0,  l.d=0.1w}{v4}
		\fmfv{l=\footnotesize$\noexpand p^{'}\rho$,     l.a=0,  l.d=0.1w}{v3}
	\end{fmfgraph*}
	}
%--------------------------------------------------------------------------
\\
&=&
\Gamma^{(s)}(p,p')
\delta_{\alpha\gamma}\delta_{\beta\rho}
+
\Gamma^{(a)}(p,p') 
\vec{\sigma}_{\alpha\gamma}\cdot\vec{\sigma}_{\beta\rho}
\,,
\nonumber
\eer
\end{fmffile}
%---------------------------------------

\vspace*{-5mm}
\noindent with
amplitudes $\Gamma^{(s)}(p,p')$ for spin-independent scattering, $\Gamma^{(a)}(p,p')$, representing
the spin-dependent ``exchange'' scattering amplitude.
The Fermion propagator in the presence of the external perturbation, $u_{\alpha\beta}(p)$, is represented by
%---------- Gamma-ph Vertex ------------

\setlength{\unitlength}{1mm}
\begin{fmffile}{fmf_G}
\input{feynman_defs.sty}
\be
G_{\alpha\beta}(p) 
= \,\,
\langle\psi_{\alpha}(p)\bar\psi_{\beta}(p)\rangle
\equiv\,\,
%--------------------- Anomalous Propagator -------------------------------
	\parbox{25mm}{
	\begin{fmfgraph*}(25,5)
		\fmfforce{(0.1w,0.5h)}{v1}
		\fmfforce{(0.9w,0.5h)}{v2}
		\fmfforce{(0.1w,0.5h)}{v11}
		\fmfforce{(0.9w,0.5h)}{v21}
		\fmfv{l=\footnotesize$\noexpand \alpha$,l.a=180,l.d=0.051w}{v1}
		\fmfv{l=\footnotesize$\noexpand \beta$, l.a=0  ,l.d=0.051w}{v2}
		\fmfv{l=\footnotesize$\noexpand {+p}$,  l.a=90, l.d=0.051w}{v11}
		\fmfv{l=\footnotesize$\noexpand {+p}$,  l.a=90, l.d=0.051w}{v21}
		\fmf{double_fermion}{v2,v1}
	\end{fmfgraph*}
	}
%--------------------------------------------------------------------------
\quad,
\ee
\end{fmffile}
%---------------------------------------

\vspace*{-5mm}
\noindent where $p=(\vp,\varepsilon_n)$ is the four-momentum, $\varepsilon_n=(2n+1)\pi T$ is the Fermion Matsubara 
energy, and $\alpha$ and $\beta$ are the initial and final state spin projections defining the Fermion propagator. 

For \He{} quasiparticles and pairs confined to a low-energy band near the Fermi surface, 
the vertex function, which varies slowly with $|\vp|$ 
in the neighborhood of the Fermi surface, can be evaluated with $\vp=p_f\hp$,
$\varepsilon_n\rightarrow 0$ and $\vp'=p_f\hp'$, $\varepsilon_n'\rightarrow 0$ within the low-energy band,
$|\varepsilon_n|, |\varepsilon_n'| \le \hbar\Omega_c$. In the same limit, we approximate the
momentum space integral as $\int\dthree{p'}(\ldots)\rightarrow
\int\frac{d\Omega_{\hp'}}{4\pi}N(0)\int d\xi_{\vp'}(\ldots)$. The resulting vertex part reduces to
functions of the relative momenta,
$A^{(s,a)}(\hp,\hp')=2N(0)\Gamma^{(s,a)}(p_f\hp,\eps=0;p_f\hp',\eps'=0)$, where $N(0)$ is the density of states
at the Fermi level and $\xi_{\vp}=v_f(|\vp|-p_f)$ is the quasi-particle excitation energy in the low-energy band
near the Fermi surface.
Rotational invariance implies that the vertex part can 
be expanded in terms of basis functions of the irreducible representations of $\orbital$, i.e.
spherical harmonics, $\{Y_{\ell,m}(\hp)|\,m=-\ell\ldots+\ell\}$, defined on the Fermi surface,
\be\label{eq-ph_scattering-amplitude}
A^{(s,a)}(\hp,\hp') 
= 
\sum_{\ell}\,A^{(s,a)}_{\ell}
\sum_{m=-\ell}^{+\ell}\,Y_{\ell,m}(\hp)\,Y_{\ell,m}^{*}(\hp')
\,,
\ee
where the sum is over relative angular momentum channels, $\ell \ge 0$.
The resulting spin independent ($\Sigma(\hp)$) and exchange ($\vec{\Sigma}(\hp)$) self-energies defined
on the low-energy bandwidth of the interaction are given by
\ber
\hspace*{-5mm}
\Sigma(\hp) 
	&=& \Sigma_{\text{ext}}(\hp)
	 +
	\int\,\frac{d\Omega_{\hp'}}{4\pi}\,A^{(s)}(\hp,\hp')\,
	T\sum_{\varepsilon_{n'}}\,\nt{'}\,
	\,g(\hp',\eps_n')
\,,
\label{eq-Scalar}
\\
\vec{\Sigma}(\hp) 
	&=& \vec{\Sigma}_{\text{ext}}(\hp)
	 +
	\int\,\frac{d\Omega_{\hp'}}{4\pi}\,A^{(a)}(\hp,\hp')\,
	T\sum_{\varepsilon_{n'}}\,\nt{'}\,
	\vec{g}(\hp',\eps_n')
\,,
\label{eq-Vector}
\eer
where $g$ and $\vec{g}$ are the scalar and spin-vector components of the quasi-classical propagator 
obtained by integration over the momentum shell $-\Omega_c\le v_f\delta p\le\Omega_c$ near the Fermi surface, 
$\int\,d\xi_{\vp}\,G_{\alpha\beta}(p)$
$\equiv$ 
$g_{\alpha\beta}(\hp,\eps_n)$ 
$=$ 
$g(\hp,\eps_n)\,\delta_{\alpha\beta}$ 
$+$ 
$\vec{g}(\hp,\eps_n)\cdot\,\vec{\sigma}_{\alpha\beta}$.
Note that the Matsubara sum, $\sum\,\nt{'}$, is restricted to $|\varepsilon_n'|\le\hbar\Omega_c$, and the self
energies vanish for the undisturbed Fermi sea.

\subsection{Particle-Particle Self Energy}

The Cooper instability results from repeated scattering of Fermion pairs with zero total momentum
(Cooper channel) that leads to the formation of bound Fermion pairs. 
Unbounded growth of the particle-particle amplitude is regulated by the formation of a new 
ground state, defined in terms of a macroscopic amplitude 
%
%---------- F propagator  ------------

\setlength{\unitlength}{1mm}
\begin{fmffile}{fmf_F}
\input{feynman_defs.sty}
\be
F_{\alpha\beta}(p) 
= \,\,
\langle\psi_{\alpha}(p)\psi_{\beta}(-p)\rangle
\equiv\,\,
%--------------------- Anomalous Propagator -------------------------------
	\parbox{25mm}{
	\begin{fmfgraph*}(25,5)
		\fmfforce{(0.1w,0.5h)}{v1}
		\fmfforce{(0.9w,0.5h)}{v2}
		\fmfforce{(0.1w,0.5h)}{v11}
		\fmfforce{(0.9w,0.5h)}{v21}
		\fmfv{l=\footnotesize$\noexpand \alpha$,l.a=180,l.d=0.051w}{v1}
		\fmfv{l=\footnotesize$\noexpand \beta$, l.a=0  ,l.d=0.051w}{v2}
		\fmfv{l=\footnotesize$\noexpand {+p}$,  l.a=90, l.d=0.051w}{v11}
		\fmfv{l=\footnotesize$\noexpand {-p}$,  l.a=90, l.d=0.051w}{v21}
		\fmf{pair_fermion}{v2,v1}
	\end{fmfgraph*}
	}
%--------------------------------------------------------------------------
\quad,
\ee
\end{fmffile}
%---------------------------------------

\vspace*{-3mm}
\noindent for a condensate of Fermion pairs with zero center of mass energy and momentum. 
The condensate and interaction in the Cooper channel also generates an associated mean-field
%
%---------- Delta ----------------------
\begin{fmffile}{fmf_Delta}
\input{feynman_defs.sty}
\setlength{\unitlength}{1mm}
\ber
\Delta_{\alpha\beta}(p) 
&\,\,=\,\,& \,\,\,\,\,\,\,\,
%--------------------- Mean Field Self Energy  -------------------------------
	\parbox{22mm}{
	\begin{fmfgraph*}(18,18)
		\fmfforce{(0.10w,0.50h)}{v2}
		\fmfforce{(0.90w,0.50h)}{v4}
		\fmfforce{(0.375w,0.50h)}{v5}
		\fmfforce{(0.675w,0.50h)}{v8}
		\fmfforce{(0.375w,0.75h)}{v6}
		\fmfforce{(0.675w,0.75h)}{v7}
		\fmfforce{(0.525w,0.625h)}{v9}
		\fmfv{d.sh=square,d.fi=0.0,d.si=0.25w,
			  l=$\hspace*{-1mm}\vspace*{-3.5mm}\noexpand \Gamma^{\mbox{\tiny pp}}$}{v9}
		\fmf{fermion}{v8,v4}
		\fmf{fermion}{v5,v2}
		\fmf{pair_fermion,right=3.5,width=1.25pt}{v7,v6}
		\fmfv{l=\footnotesize$\noexpand +p\; \alpha$,    l.a=180, l.d=0.05w}{v2}
		\fmfv{l=\footnotesize$\noexpand -p\; \beta$,     l.a=0,   l.d=0.05w}{v4}
	\end{fmfgraph*}
	}
%--------------------------------------------------------------------------
\nonumber\\
&\,\,=\,\,& 
-
T\sum_{\varepsilon_n'}\int\dthree{p'}\,
\Gamma^{\text{\footnotesize pp}}_{\alpha\beta;\gamma\rho}(p,p')\,F_{\gamma\rho}(p')
\,,
\label{eq-Delta_one-loop}
\eer
\end{fmffile}
%---------------------------------------
where 
\vspace*{-5mm}
%---------- Gamma-pp Vertex ------------
%
\setlength{\unitlength}{1mm}
\begin{fmffile}{fmf_Gamma-pp}
\input{feynman_defs.sty}
\ber
\hspace*{-5mm}
\Gamma^{\mbox{\footnotesize pp}}_{\alpha\beta;\gamma\rho}(p,p') 
&=&\hspace*{0.75cm}
%
%--------------------- Particle-Particle Vertex -------------------------------
	\parbox{22mm}{
	\begin{fmfgraph*}(18,18)
		\fmfforce{(0.01w,0.75h)}{v1}
		\fmfforce{(0.01w,0.25h)}{v2}
		\fmfforce{(0.99w,0.75h)}{v3}
		\fmfforce{(0.99w,0.25h)}{v4}
		\fmfforce{(0.375w,0.375h)}{v5}
		\fmfforce{(0.375w,0.625h)}{v6}
		\fmfforce{(0.625w,0.375h)}{v8}
		\fmfforce{(0.625w,0.625h)}{v7}
		\fmfforce{(0.50w,0.50h)}{v9}
		\fmfv{d.sh=square,d.fi=0.0,d.si=0.25w,			  
			  l=$\hspace*{-4mm}\noexpand \Gamma^{\mbox{\tiny pp}}$}{v9}
		\fmf{plain}{v5,v6}
		\fmf{plain}{v6,v7}
		\fmf{plain}{v7,v8}
		\fmf{plain}{v8,v5}
		\fmf{fermion}{v1,v6}
		\fmf{fermion}{v5,v2}
		\fmf{fermion}{v3,v7}
		\fmf{fermion}{v8,v4}
		\fmfv{l=\footnotesize$\noexpand +p\;   \alpha$,   l.a=180,  l.d=0.1w}{v2}
		\fmfv{l=\footnotesize$\noexpand +p^{'} \gamma$,   l.a=180,  l.d=0.1w}{v1}
		\fmfv{l=\footnotesize$\noexpand -p\;   \beta$,    l.a=0,    l.d=0.1w}{v4}
		\fmfv{l=\footnotesize$\noexpand -p^{'} \rho$,     l.a=0,    l.d=0.1w}{v3}
	\end{fmfgraph*}
	}
%--------------------------------------------------------------------------
\nonumber\\
&=&
\Gamma^{(0)}(p,p')
(i\sigma_y)_{\alpha\beta}(i\sigma_y)_{\gamma\rho}
\\
&+&
\Gamma^{(1)}(p,p') 
(i\vec{\sigma}\sigma_y)_{\alpha\beta}\cdot(i\sigma_y\vec{\sigma})_{\gamma\rho}
\,,
\eer
\end{fmffile}
%---------------------------------------

\vspace*{-5mm}
\noindent is the four-Fermion vertex that is irreducible in the particle-particle channel, expressed
in terms of the spin-singlet ($S=0$), even-parity and spin-triplet ($S=1$), odd-parity pairing
interactions, $\Gamma^{(0)}(p,p')$ and $\Gamma^{(1)}(p,p')$, respectively.
Thus, the pairing self energy separates into singlet and triplet components
\be
\Delta_{\alpha\beta}(p) 
=
d(p)\,(i\sigma_y)_{\alpha\beta} 
+ 
\vec{d}(p)\cdot(i\vec{\sigma}\sigma_y)_{\alpha\beta}
\,.
\ee
Fermion pairs with binding energy $|\Delta|<\Omega_c$ are confined to a low-energy band
near the Fermi surface, $|\varepsilon|\le\hbar\Omega_c\ll E_f$, and a shell in momentum space,
$\delta p \le \hbar\Omega_c/v_f$. Thus, the particle-particle irreducible vertex, which varies
slowly on with $|\vp|$ in the neighborhood of the Fermi surface, can also be evaluated with 
$\vp=p_f\hp$, $\varepsilon_n\rightarrow 0$ and $\vp'=p_f\hp'$, $\varepsilon_n'\rightarrow 0$.
Thus, 
$\Gamma^{pp}$ 
reduces to even- and odd-parity functions of the relative momenta,
$V^{(S)}(\hp,\hp')=2N(0)\Gamma^{(S)}(p_f\hp,\eps=0;p_f\hp',\eps'=0)$,
and rotational invariance of the normal-state Fermionic vacuum implies
\ber
V^{{\binom{0}{1}}}(\hp,\hp') 
&=& 
-
\sum^{\binom{\text{even}}{\text{odd}}}_{\ell}\,v_{\ell}
\sum_{m=-\ell}^{+\ell}\,Y_{\ell,m}(\hp)\,Y_{\ell,m}^{*}(\hp')
\nonumber\\
&=& 
-
\sum^{\binom{\text{even}}{\text{odd}}}_{\ell}\,(2\ell+1)\,v_{\ell}\,P_{\ell}(\hp\cdot\hp')
\,,
\label{eq-pairing_interactions}
\eer
where the sum is over all even (odd) orbital angular momentum channels, $\ell \ge 0$, for spin-singlet
(spin-triplet) pair scattering, and $-v_{\ell}$ is the pairing interaction (``coupling constant'')
in the orbital angular momentum channel $\ell$.\footnote{$v_{\ell}>0$ corresponds to an 
{\sl attractive} interaction in channel $\ell$.}
The singlet ($d(\hp)$) and triplet ($\vec{d}(\hp)$) self-energies are given by
\ber
d(\hp) 
&=& 
-	\int\,\frac{d\Omega_{\hp'}}{4\pi}\,V^{(0)}(\hp,\hp')\,\,
	T\sum_{\varepsilon_{n'}}\,\nt{'}\,
	f(\hp',\eps_n')
\,,
\label{eq-singlet_d}
\\
\vec{d}(\hp) 
&=& 
-	\int\,\frac{d\Omega_{\hp'}}{4\pi}\,V^{(1)}(\hp,\hp')\,\,
	T\sum_{\varepsilon_{n'}}\,\nt{'}\,
	\vec{f}(\hp',\eps_n')
\,,
\label{eq-triplet_d}
\eer
where 
$f_{\alpha\beta}(\hp,\eps_n)$ 
$\equiv$ 
$\int\,d\xi_{\vp}\,F_{\alpha\beta}(p)$
$=$ 
$f(\hp,\eps_n)\,(i\sigma_y)_{\alpha\beta}$ 
$+$ 
$\vec{f}(\hp,\eps_n)\cdot\,(i\vec{\sigma}\sigma_y)_{\alpha\beta}$ 
is the quasi-classical pair propagator expressed in terms of the anomalous singlet and triplet
components, $f$ and $\vec{f}$.

The breaking of $\point{U(1)}{\ns}$ symmetry by pair condensation implies mixing of normal-state 
particle- and hole states. Particle-hole coherence is accommodated by introducing Nambu spinors,
$\Psi=\left(\psi_{\uparrow}\,,\,\psi_{\downarrow}\,,\,
	  \psi^{\dag}_{\uparrow}\,,\,\psi^{\dag}_{\downarrow}\right)$,
or equivalently by a $4\times 4$ Nambu matrix propagator in the combined particle-hole and spin space.
In the quasi-classical limit the Nambu propagator is represented by the diagonal and off-diagonal 
quasiclassical propagators, $\hg$ and $\hf$, and their conjugates, $\hg'$ and $\hf'$,
\be
\whg = 
\begin{pmatrix}	
	g + \vec{g}\cdot\vec{\vsigma} 
& 	f \, i\sigma_y + \vec{f}\cdot i\vec{\vsigma}\sigma_y 
\\	f' \, i\sigma_y + \vec{f'} \cdot i\sigma_y\vec{\vsigma} 
&	g' - \vec{g'}\cdot \sigma_y\vec{\vsigma}\sigma_y
\end{pmatrix}
\,,
\ee
where $g$ ($\vec{g}$) is the spin scalar (vector) component of the Fermion propagator, while $f$
($\vec{f}$) is the spin singlet (triplet) component of the anomalous pair propagator. The lower row
of the Nambu matrix represents the conjugate propagators, $\hgbar$ and $\hfbar$, which are related to
$\hg$ and $\hf$ by the combination of Fermion anti-symmetry and particle-hole conjugation symmetries
(c.f. App. \ref{sec-appendix-symmetries}). 
Similarly, the quasiparticle and pairing self-energies are organized into a $4\times 4$ Nambu matrix,
\be\label{eq-Self-Energy}
\whSigma
= 
\begin{pmatrix}	
	\Sigma + \vec{\Sigma}\cdot\vec{\vsigma} 
& 	d \, i\sigma_y + \vec{d} \cdot i\vec{\vsigma}\sigma_y 
\\
	d' \, i\sigma_y + \vec{d'} \cdot i\sigma_y\vec{\vsigma} 
&	\Sigma' - \vec{\Sigma}'\cdot \sigma_y\vec{\vsigma}\sigma_y
\end{pmatrix}
\,,
\ee
with the corresponding symmetry relations connecting the conjugate self-energies to 
$\Sigma$, $\vec{\Sigma}$, $d$ and $\vec{d}$.
This doubling of the Fermionic and Bosonic degrees of freedom, which is forced by the breaking of 
global $\point{U(1)}{\ns}$ symmetry, is the origin of the doublets of Bosonic modes labeled by parity under 
charge conjugation, $\charge=\pm 1$, in BCS-type theories.

\vspace*{-3mm}
\section{Eilenbergers' Equations}
\vspace*{-3mm}

The quasiparticle and anomalous pair propagators and self-energies, organized into $4\times 4$ Nambu 
matrices, obey Gorkov's equations.\cite{gor58} Eilenberger transformed Gorkov's equations into a matrix 
transport-type equation for the quasiclassical propagator and self-energy,\cite{eil68} 
\be\label{eq-Eilenberger_Equilibrium}
\commutator{i\varepsilon_n\whtauz - \whSigma(\hp,\vR)}{\whg} + i\hbar\vv_{\hp}\cdot\grad\whg = 0
\,.
\ee
In contrast to Gorkov's equation, which is a second-order differential equation with a unit source term 
originating from the Fermion anti-commutation relations, Eilenberger's equation is a homogeneous, 
first-order 
differential equation describing the evolution of the quasiclassical propagator along classical trajectories 
defined by the Fermi velocity, $\vv_{\hp}=v_f\hp$. 
The form of Eilenberger's equation in Eq. \ref{eq-Eilenberger_Equilibrium} governs the \emph{equilibrium} 
propagator, including inhomogeneous states described by an external potential or a spatially varying 
mean pairing self-energy, $\whDelta(\hp,\vR)$, but must be supplemented by the 
normalization condition,\cite{eil68}
\be\label{eq-normalization-zeroth}
\whg(\hp,\varepsilon_n;\vR)^2 = -\pi^2\,\tone
\,,
\ee
which restores the constraint on the spectral weight implied by the source term in Gorkov's equation.
For the spatially homogeneous ground-state of superfluid \Heb{} Eilenberger's equation reduces to
\be
\commutator{i\varepsilon_n\whtauz - \whSigma(\hp)}{\whg_0} = 0
\,,
\ee
and the homogeneous self-energy, $\whSigma\equiv\whDelta(\hp)$, is defined by the mean-field pairing 
self-energy for the \He\ ground state,
\be\label{eq-Triplet_Order_Parameter}
\whDelta(\hp) = 
\begin{pmatrix}		0		&	\vec{\Delta}(\hp)\cdot i\vec{\sigma}\sigma_y 
\\
	\vec{\Delta}(\hp)\cdot i\sigma_y\vec{\sigma} &	0	
\end{pmatrix}
\,,
\ee
where $\vec{\Delta}(\hp) = \Delta\,\hp$ is the $J=0$ BW order parameter. Here and after we denote the 
equilibrium spin-triplet order parameter as $\vec{\Delta}$ and reserve $\vec{d}$ for the non-equilibrium 
fluctuations of the spin-triplet order parameter.
The $4\times 4$ matrix order parameter for the BW state satisfies, 
$\whDelta(\hp)\whDelta(\hp) = -|\Delta|^2\,\tone$. 
Thus, the equilibrium propagator for the BW state is given by
\be\label{eq-QC_propagator-BW_equilibrium}
\whg_0(\hp,\varepsilon_n) = -\pi\frac{i\varepsilon_n\whtauz - \whDelta(\hp)}
				     {\sqrt{\varepsilon_n^2 + |\Delta|^2}}
\,.
\ee
Note that the diagonal component of $\whg_0$ is odd in frequency. This implies that the diagonal (Fermionic) 
self-energies, $\Sigma(\hp)$ and $\vec\Sigma(\hp)$ evaluated with Eq. \ref{eq-QC_propagator-BW_equilibrium}, 
vanish in equilibrium. 
However, if the ground state is perturbed, e.g. by a Bosonic fluctuation of the Cooper pair condensate, the 
Fermionic self energy, in general, no longer vanishes.

In equilibrium, the anomalous self-energy reduces to the self-consistency equation (``gap equation'') 
for the spin-triplet order parameter,
\be\label{eq-B-phase_Order_Parameter}
\vec{\Delta}(\hp) = -\frac{\pi}{\beta}\sum_{\varepsilon_n}{'}\int\frac{d\Omega_{\hp}}{4\pi}\,
                  V^{(1)}(\hp,\hp')\,\frac{\vec{\Delta}(\hp')}{\sqrt{\varepsilon_n^2+|\vec{\Delta}(\hp')|^2}}
\,.
\ee
The linearized gap equation defines the instability temperatures for Cooper pairing with orbital angular 
momentum $\ell$,
\be\label{eq-Linearized_Gap-Equation}
\frac{1}{v_{\ell}} = \pi T_{c_{\ell}}\sum_{\varepsilon_n}{'}\frac{1}{|\varepsilon_n|} \equiv K(T_{c_{\ell}}) 
\,,
\ee
for attractive interactions $v_{\ell}>0$.
The function $K(T)$ is a digamma function of argument, $\hbar\Omega_c/2\pi T \gg 1$, in which case
\be\label{eq-K_diGamma}
K(T) \equiv \pi T\sum_{\varepsilon_n}{'}\,\frac{1}{|\varepsilon_n|} 
\simeq
\ln\left(\frac{2e^{\gamma_{\text{E}}}}{\pi}\frac{\hbar\Omega_c}{T}\right) 
\,,
\ee
where $\gamma_{\text{E}}\simeq 0.57721$ is Euler's constant. This function plays a central role in regulating 
the log-divergence of frequency sums in the Cooper channel. 
In \He\ the p-wave pairing channel is the dominant attractive channel; the f-wave channel is also attractive, 
but sub-dominant, i.e. $0 < T_{c_3} < T_{c_1} \equiv T_c$.

The anomalous self-energy in the p-wave channel also determines the mass (gap), $\mF=\Delta$, of 
Fermionic excitations of the 
Balian-Werthamer phase. In particular the p-wave projection of Eq. \ref{eq-B-phase_Order_Parameter} reduces 
to the BCS gap equation,
\be\label{eq-BCS_gap-regulated}
\ln(T/T_c) = 2\pi T\sum_{n\ge 0}^{\infty}
             \left(
	     \frac{1}{\sqrt{\varepsilon_n^2 + \Delta^2}}
	     - 
	     \frac{1}{\varepsilon_n} 
	     \right)
\,.
\ee
Note that both the pairing interaction, $v_1$ and cutoff, $\Omega_c$, in 
Eq. \ref{eq-B-phase_Order_Parameter} have been 
eliminated in favor of the transition temperature by regulating the log-divergent sum 
using Eq. \ref{eq-K_diGamma} and 
the linearized gap equation for $T_c$, Eq. \ref{eq-Linearized_Gap-Equation}.

The Balian-Werthamer state, which has an isotropic gap in the Fermionic spectrum, is maximally effective in using 
states near the Fermi surface for pair condensation. As a result the B-phase is stable down to $T=0$ in spite 
of the attractive f-wave pairing interaction.\cite{sau86} 
Nevertheless, sub-dominant f-wave pairing plays an important role in the Bosonic excitation spectrum of the 
B-phase. In particular, p-wave, spin-triplet Higgs modes with $J=2$ polarize the B-phase vacuum. The $J=2$ 
polarization couples to f-wave, spin-triplet Cooper pair fluctuations with $J=2$, leading to 
mass corrections to the $J^{\charge}=2^{\pm}$ Higgs modes. In the following we derive the 
dynamical equations for the Bosonic modes including the polarization terms from the f-wave pairing channel, and
self-energy corrections from the Landau channel.

\vspace*{-3mm}
\section{Dynamical Equations}
\vspace*{-3mm}

In order to describe the non-equilibrium response, or fluctuations relative to homogeneous equilibrium,
we must generalize the low-energy quasiparticle and Cooper pair propagators to functions of two time 
($\tau_1,\tau_2$), or frequency ($\varepsilon_{n_1},\varepsilon_{n_2}$), variables. Specifically, we must 
include the dependence on the global time coordinate, $\Upsilon=(\tau_1 + \tau_2)/2$, or equivalently the 
total Matsubara energy $\omega_m$, in addition to the relative time difference, $\tau_1-\tau_2$, or
corresponding Fermion Matsubara energy, $\varepsilon_n$.
Thus, the $\xi_{\vp}$-integrated quasiclassical propagator generalizes to 
$\whg(\hat{p},\varepsilon_n)\rightarrow \whg(\hat{p},\varepsilon_n;\vq,\omega_m)$, where $\vq$ is the total 
momentum, or wave vector for a Fourier mode associated with the center of mass coordinate, $\vR$.

The space-time dynamics of the coupled system of Fermionic and Bosonic excitations of the broken
symmetry ground state is encoded in the Keldysh propagator,\cite{kel65} which is obtained here by analytic
continuation to the real energy axes, e.g. $i\varepsilon_n\rightarrow\varepsilon + i0^{+}$ followed by
$i\omega_m\rightarrow\omega + i0^{+}$.
Thus, 
\be\label{eq-analytic_continuation}
\frac{1}{\beta}\sum_{\varepsilon_n}\,\whg(\varepsilon_n;\omega_m)
	\xrightarrow[i\omega_m\rightarrow\omega+i0^{+}]{}
\int_{-\infty}^{+\infty}\,\frac{d\varepsilon}{4\pi i}\,\whg^{K}(\varepsilon;\omega)
\,,
\ee
where $\whg^{K}(\hp,\varepsilon;\vq,\omega)$ is the real-energy, and frequency dependent Keldysh
propagator. The Keldysh propagator determines the response to any space-time dependent excitation. 
For example the particle current is given by,
\be
\vJ
= 
N(0)\int\frac{d\Omega_{\hp}}{4\pi}\,
   \int\frac{d\varepsilon}{4\pi i}\,(\vv_{\vp})
	\Tr{\whtauz\whg^{K}(\hp,\varepsilon;\vq,\omega)}
\,.
\ee

The off-diagonal Nambu components of the Kelysh propagator determine the Bosonic modes of
the interacting Fermionic and Bosonic system. The spin-triplet Bosonic excitations are obtained 
from the anomalous triplet propagator, $\vec{f}^{K}$, and the self-consistent solution for the anomalous 
self-energy obtained by analytic continuation of Eq. \ref{eq-triplet_d},
\be
\vec{d}(\hp;\vq,\omega)
= 
-
%\nicefrac{1}{2}
\int\frac{d\Omega_{\hp}}{4\pi}\, V^{(1)}(\hp,\hp')\, 
\int\frac{d\varepsilon}{4\pi i}\, \vec{f}^{K}(\hp,\varepsilon;\vq,\omega)
\,.
\ee

To calculate the Keldysh propagator, $\whg^K$, we generalize Eilenberger's transport equation for the two-time/frequency 
non-equilibrium Matsubara propagator, 
\be\label{eq-Eilenberger_non-equilibrium}
\left[i\varepsilon\whtauz - \whSigma\right]\circ\whg - \whg\circ\left[i\varepsilon\whtauz-\whSigma\right] 
+ i\vv_{\hp}\cdot\grad\whg = 0
\,,
\ee
where the 
$A\circ B(\varepsilon_{n_1},\varepsilon_{n_2})\equiv\frac{1}{\beta}\sum_{n_3}\,
 A(\varepsilon_{n_1},\varepsilon_{n_3})\,B(\varepsilon_{n_3},\varepsilon_{n_2})$ 
is a convolution in Matsubara energies.
For the two-frequency, non-equilibrium propagator the normalization condition is also a convolution
product in Matsubara frequencies,
\be\label{eq-normalization}
\whg\circ\whg %[\varepsilon_{n_1},\varepsilon_{n_2}]
\equiv
\frac{1}{\beta}\sum_{\varepsilon_{n_3}}
\whg(\varepsilon_{n_1},\varepsilon_{n_3})
\whg(\varepsilon_{n_3},\varepsilon_{n_2})
=
-\pi^2\,\beta\delta_{\varepsilon_{n_1},\varepsilon_{n_2}}\,\tone
\,.
\ee
If we express the full propagator as a correction to the equilibrium propagator 
(Eq. \ref{eq-QC_propagator-BW_equilibrium}),
\be
\hspace*{-2mm}
\whg(\hp,\vq;\varepsilon_{n_1},\varepsilon_{n_2})
=
\whg_{0}(\hp,\varepsilon_{n_1}) \,\beta\delta_{\varepsilon_{n_1},\varepsilon_{n_2}}
+
\delta\whg(\hp,\vq;\varepsilon_{n_1},\varepsilon_{n_2})
\,,
\ee
then to linear order in $\delta\whg$ the normalization condition for the correction to the propagator becomes
after setting
$\varepsilon_{n_1} = \varepsilon_{n}+\omega_{m}$,
$\varepsilon_{n_2} = \varepsilon_{n}$, and
$\delta\whg(\hp,\vq;\varepsilon_{n_1},\varepsilon_{n_2})
\equiv
\delta\whg(\hp,\vq;\varepsilon_{n},\omega_{m})$
\be
%\whg_{0}(\hp,\varepsilon_n+\omega_m)\delta\whg(\hp,\vq;\varepsilon_n,\omega_m)
\whg_{0}(\varepsilon_n+\omega_m)\delta\whg(\varepsilon_n,\omega_m)
+
%\delta\whg(\hp,\vq;\varepsilon_n,\omega_m)\whg_{0}(\hp,\varepsilon_n)
\delta\whg(\varepsilon_n,\omega_m)\whg_{0}(\varepsilon_n)
\,=\,0
\,.
\label{eq-normalization-first}
\ee

The Bosonic modes of the interacting Fermi superfluid are obtained from the linearized dynamical equations 
for the fluctuations of the anomalous self energy, $\delta\hDelta = \hDelta - \hDelta_0$, where 
the equilibrium self-energy, $\hDelta_0$, is defined by off-diagonal mean-field pairing 
self-energy for the \He\ ground state (Eqs. \ref{eq-Triplet_Order_Parameter} and \ref{eq-B-phase_Order_Parameter}).
These fluctuations are coupled to fluctuations of the Fermionic self-energy, $\delta\hSigma$.   
The coupled dynamical equations for the components of $\delta\whSigma(\hp;\vq,\omega)$ are obtained solving the 
non-equilibrium Eilenberger equation, Eq. \ref{eq-Eilenberger_non-equilibrium}, for $\whg$ to linear order in the 
self-energy fluctuations, $\delta\whSigma$.
The linearized non-equilibrium Eileberger equation becomes, 
\ber
&&
\hspace*{-5mm}
\left\{i(\varepsilon_n+\omega_m)\whtauz - \whDelta(\hp)\right\}\,\delta\whg-\delta\whg\,
\left\{i\varepsilon_n\whtauz - \whDelta(\hp)\right\}
- 
\vv_{\hp}\cdot\vq\,\,\delta\whg
\nonumber\\
&&
\hspace*{5mm}
+\,\whg_0(\hp,\varepsilon_n+\omega_m)\,\delta\whSigma - \delta\whSigma\,\whg_0(\hp,\varepsilon_n) 
= 0
\,.
\label{eq-Eilenberger-linearized}
\eer
%--------------------- 
%\begin{widetext}
%\ber
%&&\left[i(\varepsilon_n + \omega_m)\whtauz - \whDelta(\hp)\right]\,\delta\whg(\hp,\varepsilon_n;\vq,\omega_m)
%- 
%\delta\whg(\hp,\varepsilon_n;\vq,\omega_m)\,\left[i\varepsilon_n\whtauz - \whDelta(\hp)\right]
%\nonumber\\
%&&\qquad\qquad +
%\whg_0(\hp,\varepsilon_n+\omega_m)\,\delta\whSigma(\hp,\varepsilon_n;\vq,\omega_m)
%- 
%\delta\whSigma(\hp,\varepsilon_n;\vq,\omega_m)\,\whg_0(\hp,\varepsilon_n)
%-
%\vv_{\hp}\cdot\vq\,\,\delta\whg(\hp,\varepsilon_n;\vq,\omega_m)
%= 0
%\,.
%\eer
%\end{widetext}
%--------------------- 
%
The normalization conditions, Eqs. \ref{eq-normalization-zeroth} and \ref{eq-normalization-first},
combined with Eq. \ref{eq-QC_propagator-BW_equilibrium}, provides a direct method of inverting 
Eq. \ref{eq-Eilenberger-linearized} for the non-equilibrium quasiclassical propagator,
\begin{widetext}
\be\label{eq-QC_Linear_Response}
\delta\whg 
=
\left(\frac{-1}{\pi^2D_{+}^2+(\vv_{\hp}\cdot\vq)^2}\right)
\left[D_{+}\,
	\left\{\whg_{0}(\varepsilon_n+\omega_m)\,\delta\whSigma\,\whg_{0}(\varepsilon_n) 
	+ \pi^2\,\delta\whSigma
	\right\}
	+
	\vv_{\hp}\cdot\vq\,
	\left\{\delta\whSigma\,\whg_{0}(\varepsilon_n) - \whg_{0}(\varepsilon_n+\omega_m)\,\delta\whSigma\right\}
\right]
\,.
\ee
\end{widetext}
where $D_{+}(\varepsilon_n,\omega_m)=D(\varepsilon_n+\omega_m)+D(\varepsilon_n)$, and 
\be
D(\varepsilon_n)\equiv\frac{-1}{\pi}\sqrt{\varepsilon_n^2 + |\Delta|^2}
\,,
\ee
is the denominator of the equilibrium propagator.

To calculate the mass spectrum of the Bosonic modes we need only the $\vq=0$ propagators, in which case
\be\label{eq-QC_Linear_Response_q=0}
\delta\whg = \frac{-1}{\pi^2}\frac{1}{D_{+}}\,
	     \left\{\whg_{0}(\varepsilon_n+\omega_m)\,\delta\whSigma\,\whg_{0}(\varepsilon_n) 
	           + \pi^2\,\delta\whSigma
	     \right\}
\,.
\ee
In zero magnetic field, spin-singlet Bosonic fluctuations, if they exist, do not couple to spin-triplet Bosonic 
fluctuations.  However, we must retain fluctuations of the Fermionic self-energy, thus the form of the fluctuation 
self-energy becomes,
\be\label{eq-Dynamical_Self-Energy}
\delta\whSig
= 
\begin{pmatrix}	
	\Sigma + \vec{\Sigma}\cdot\vec{\vsigma} 
& 	\vec{d} \cdot i\vec{\vsigma}\sigma_y
\\      \vec{d'} \cdot i\sigma_y\vec{\vsigma}
&	\Sigma' - \vec{\Sigma'}\cdot \sigma_y\vec{\vsigma}\sigma_y
\end{pmatrix}
\,.
\ee
where the conjugate spin-triplet order parameter amplitudes are related 
by $\vec{d}'(\hp;\vq,\omega_m)=\vec{d}(\hp;-\vq,-\omega_m)^*$ (see App. \ref{sec-appendix}).
The linear combinations,
\be
\vec{d}^{(\pm)}(\hp;\vq,\omega_m) \equiv
%\nicefrac{1}{2}\left[
\vec{d}(\hp;\vq,\omega_m) \pm \vec{d}'(\hp;\vq,\omega_m)
%\right]
\,,
\ee
have charge conjugation parities, $\charge=\pm 1$; the dynamical equations for Bosonic modes then 
separate into charge conjugation doublets with opposite parity.
The Bosonic modes of Cooper pairs also couple to the fluctuations of the Fermionic self energy, in both 
the spin scalar and vector channels,
\ber
\Sigma^{(\pm)}(\hp;\vq,\omega_m) 
\equiv
\Sigma(\hp;\vq,\omega_m) \pm \Sigma'(\hp;\vq,\omega_m)
\,,
\\
\vec\Sigma^{(\pm)}(\hp;\vq,\omega_m) 
\equiv
\vec\Sigma(\hp;\vq,\omega_m) \pm \vec\Sigma'(\hp;\vq,\omega_m)
\,.
\eer
Note that the exchange and conjugation symmetry relations for the diagnoal self-energies, 
Eqs. \ref{eq-exchange_symmetry-scalar}-\ref{eq-exchange_symmetry-vector} and 
Eqs. \ref{eq-conjugation_symmetry-scalar}-\ref{eq-conjugation_symmetry-vector},
imply that the Fermionic self-energies, $\Sigma^{(\pm)}$ and $\vec\Sigma^{(\pm)}$ 
are also even (odd) with respect to charge conjugation parity, $\charge=\pm 1$.

The dynamical equations for the spin-triplet Bosonic modes are obtained from the
off-diagonal {\it and} diagonal components of $\delta\whg$ in Eq. \ref{eq-QC_Linear_Response}, the self-consistency 
equations for the leading order mean-field self-energies, Eqs. \ref{eq-Scalar}, \ref{eq-Vector} 
and \ref{eq-triplet_d}.
Two response functions are obtained from the propagator in Eq. \ref{eq-QC_Linear_Response_q=0} that determine the 
Bosonic and Fermionic self-energies,
\ber
\gamma(i\omega_m) 
	&=& 
	-\frac{1}{\beta}\sum_{\varepsilon_n}
	\Big[
	\frac{1}{D(\varepsilon_n)}+\frac{1}{D(\varepsilon_n+\omega_m)}
	\Big]
\,,
\label{eq-gamma}
\\
\lambda(i\omega_m)
	&=& 
	\frac{2}{\pi^2\beta}\sum_{\varepsilon_n}
	\frac{|\Delta^2|}{D_{+}(\varepsilon_n,\omega_m)D(\varepsilon_n)D(\varepsilon_n+\omega_m)}
\,.
\label{eq-lambda}
\eer
The Matsubara sum defining $\gamma(i\omega_m)$ is log-divergent, regulated by the cutoff $\Omega_c$. The 
frequency dependence of $\gamma$ can be neglected, since it gives a negligible correction of 
order $(\omega_m/\Omega_c)^2\ll 1$.
Thus,
\be
\onehalf\gamma	= \frac{\pi}{\beta}\sum_{\varepsilon_n}{'}\,\frac{1}{\sqrt{\varepsilon_n^2 + |\Delta|^2}}
		= \frac{1}{v_1} 
\,,
\ee
where the latter equality follows from the equilibrium gap equations, 
Eqs. \ref{eq-Linearized_Gap-Equation} - \ref{eq-BCS_gap-regulated}.
The function $\lambda(i\omega_m)$ is defined by a convergent Matsubara sum. 
Analytic continuation to real frequencies of 
Eq. \ref{eq-lambda} in the manner of Eq. \ref{eq-analytic_continuation} yields,
\be
\lambda(\omega) \equiv |\Delta|^2\,\bar\lambda(\omega)
= 
|\Delta|^2\int_{|\Delta|}^{\infty}\,
                           \frac{d\varepsilon}{\sqrt{\varepsilon^2-|\Delta|^2}}
			   \frac{\tanh\left(\frac{\beta\varepsilon}{2}\right)}{\varepsilon^2-\omega^2/4}
\,,
\ee
which is the Tsuneto function with $\omega\rightarrow\omega+i0^{+}$ defining the retarded (causal) 
response.\cite{tsu60}
For $|\omega| < 2|\Delta|$, $\lambda(\omega)$ is real and defines the non-resonant frequency response of the 
condensate, while for $|\omega| > 2|\Delta|$, $\mbox{Im}\lambda(\omega)\ne 0$ is the spectral density 
of unbound Fermion pairs. 
In the $T=0$ limit, with $x=\omega/2|\Delta|$,
\be\label{eq-lambda_T=0}
\lambda(\omega) =
\left\{
\begin{matrix}
\displaystyle
\frac{\sin^{-1}(x)}{x\sqrt{1 - x^2}}
\,\, ,
& 
|x| < 1
\cr
&
\cr
\displaystyle
\frac{1}{2x\sqrt{x^2-1}}
\left[
\ln\left\vert\frac{\sqrt{x^2-1}-x}{\sqrt{x^2-1}+x}\right\vert + i \pi \sgn(x)
\right]
\ns,
& 
\ns
|x| > 1
\end{matrix}
\right.
\ee

%\be\label{eq-lambda_T=0}
%\lambda(\omega) = 
%\left\{
%\begin{matrix}
%\displaystyle
%\frac{4|\Delta|^2}{\omega}
%\frac{\arcsin
%\left(\displaystyle
%{\omega}/{2|\Delta|}
%\right)
%}
%     {\sqrt{4|\Delta|^2 - \omega^2}}
%\,,
%&
%\omega < 2|\Delta|
%\,,
%\cr
%\, & \,
%\cr
%\displaystyle
%%
%\frac{\pi}{2}
%\left(\frac{4|\Delta|^2}{\omega}\right)
%\frac{1}{\sqrt{\omega^2 - 4|\Delta|^2}}
%\,,
%&
%\omega > 2|\Delta|
%%
%\,.
%\end{matrix}
%\right.
%\ee

\begin{widetext}

Thus, analytic continuation to real frequencies for the $\vq=0$ limit leads to the following dynamical equations 
for the spin-triplet Bosonic modes of the B-phase ground state,\cite{sau81,mck90,mck90a,moo93}

%
% Spin-Triplet Components of the Dynamical Pairing Self-Energy for Real, Unitary Spin-Triplet States
% Adapted from Dynamical_Equations-Triplet_BW-q0.tex
%
\ber
\vec{d}^{(-)}(\hp;\omega) &=& - \int\dangle{p'}\,V^{(1)}(\hp,\hp')\,
	\Big\{
	\left[\onehalf\gamma + \onefourth(\omega^2 - 4|\Delta|^2)\bar\lambda(\omega)\right]\,\vec{d}^{(-)}(\hp';\omega) 
	 +
	\bar\lambda(\omega)\vec\Delta(\hp')\,(\vec\Delta(\hp')\cdot\vec{d}^{(-)}(\hp';\omega))
\nonumber\\
	&&\hspace*{30mm}
	 - \onehalf\omega\,\bar\lambda(\omega)\,\vec\Delta(\hp')\,\Sigma^{(+)}(\hp';\omega)
	\Big\}
\,,
\label{eq-triplet-odd_parity-q0}
\\
\quad
\nonumber\\
\vec{d}^{(+)}(\hp;\omega) &=& - \int\dangle{p'}\,V^{(1)}(\hp,\hp')\,
			\Big\{
			\left[\onehalf\gamma + \onefourth\omega^2\bar\lambda(\omega)\right]\,\vec{d}^{(+)}(\hp';\omega) 
			 -
			\bar\lambda(\omega)\vec\Delta(\hp')(\vec\Delta(\hp')\cdot\vec{d}^{(+)}(\hp';\omega))
\nonumber\\
			&&\hspace*{30mm}
			 + \nicefrac{i}{2}\omega\,\bar\lambda(\omega)\,\vec\Delta(\hp')\times\vec\Sigma^{(+)}(\hp';\omega)
			\Big\}
\,,
\label{eq-triplet-even_parity-q0}
\eer
\end{widetext}
Note that the equations of motion for the Bosonic fluctuations of the order parameter couple to the 
Fermionic self-energies {\it linearly} in the frequency $\omega$, and that only the even orbital parity 
Fermionic fluctuations contribute in the $\vq=0$ limit.

For the moment we omit pairing fluctuations in higher angular 
momentum channels, i.e. set $v_{\ell}=0$ for $\ell\ge 3$. We then expand the spin-triplet order parameter 
amplitudes, $\vec{d}^{(\pm)}(\hp)$, in terms of the p-wave basis, 
$d^{(\pm)}_{\alpha}(\hp) = \cD^{(\pm)}_{\alpha i}\hp_i$, where $\cD^{(\pm)}_{\alpha i}$ is equivalent 
to the bi-vector representation of the order parameter discussed in the 
context of the TDGL theory for the Bosonic modes.
For the B-phase ground state with total angular momentum $J=0$, i.e. $\vec\Delta(\hp)=\Delta\hp$, 
or equivalently, 
$A_{\alpha i}=\Delta/\sqrt{3}\,\delta_{\alpha i}$, Eqs. \ref{eq-triplet-odd_parity-q0} 
and \ref{eq-triplet-even_parity-q0}
can be solved by expanding the pairing fluctuations in spherical tensors that define 
bases for the representations of the 
residual symmetry group of the B-phase, $H=\point{SO(3)}{J}$, with total angular momentum $J=0,1,2$,
\be\label{eq-p-wave-triplet_fluctuations}
\cD^{(\pm)}_{\alpha i} = \sum_{J=0,1,2}\sum_{m=-J,+J}\,D^{(\pm)}_{J,m}\,t^{(J,m)}_{\alpha i}
\,.
\ee
Note that time-dependent fluctuations of the Fermionic self-energy, e.g. $\omega\Sigma^{(+)}(\hp;\omega)$, 
appear as ``source'' terms in the equations of motion for the order parameter collective modes. 

\vspace*{-3mm}
\subsection{Nambu-Goldstone and Higgs Modes with $\charge=-1$}\label{sec-J-Modes}
\vspace*{-3mm}

In the case of the modes with parity $\charge=-1$ we can express
\be
\Sigma^{(+)}(\hp;\omega) = \sum_J^{\text{even}}\sum_m\,\Sigma^{(+)}_{J,m}(\omega)\,\hp_i\,t^{(J,m)}_{ij}\,\hp_j
\,.
\label{eq-Sigma+Scalar_Jm}
\ee
Note that only self-energy fluctuations of even $J$ couple to the Bosonic modes with $\charge =-1$.
Equation \ref{eq-triplet-odd_parity-q0} then decouples into the dynamical equations for Bosonic mode 
amplitudes with total angular momentum $J$. In particular, the equation for dynamical fluctuations 
with $J^{\charge}=0^{-}$ is given by
\be\label{eq-J=0-}
\omega^2\,D^{(-)}_{0,0} =  2\,|\Delta|\,\omega\,\Sigma^{(+)}_{0,0}
\,.
\ee
In the simplest case the $J=0$ contribution to the Fermionic self-energy represents a fluctuation in the chemical 
potential, i.e. $\Sigma_{0,0}^{(+)}(\omega) = 2\delta\mu(\omega)$, and as discussed earlier the pairing fluctuation 
with $J^{\charge}=0^{-}$ represents time-dependent fluctuations of the phase of the B-phase ground state, i.e. 
$D^{(-)}_{0,0}=2i|\Delta|\,\vartheta(\omega)$. This is the massless Anderson-Bogoliubov mode, which in the 
time domain for $\vq=0$ obeys the Josephson phase relation, $\hbar\partial_t\vartheta = -2\delta\mu$. 
As we show below this result is un-renormalized by interactions between Fermions in either particle-hole or 
particle-particle channels.

Projecting out the pairing fluctuations with $J^{\charge}=1^{-}$ from Eq. \ref{eq-triplet-odd_parity-q0} yields,
\be\label{eq-J=1-}
\left(\omega^2 - 4|\Delta|^2\right)\,D^{(-)}_{1,m} = 0
\,.
\ee
This is a quite remarkable result: the $J^{\charge}=1^{-}$ pairing fluctuations do not couple to fluctuations 
in the Fermion self-energy. Furthermore, neither d-wave, spin-singlet, nor f-wave spin-triplet, pairing 
fluctuations couple the $J^{\charge}=1^{-}$ modes, which implies that the mass of $J^{\charge}=1^{-}$ Higgs modes, 
$M_{1,-}=2\Delta$, is un-renormalized by interactions to leading order in the expansion.

By contrast the $J^{\charge}=2^{-}$ modes obey the following dynamical equations
\be\label{eq-J=2-}
\Big[\omega^2 - \frac{12}{5}|\Delta|^2\Big]\,D^{(-)}_{2,m} =  \frac{4}{5}\,|\Delta|\,\omega\,\Sigma^{(+)}_{2,m}
\,.
\ee
In the absence of Fermion interactions in the particle-hole channel $\Sigma^{(+)}_{2,m}(\omega)$ 
represents {\sl external} stress fluctuations, $u^{(+)}_{2,m}(\omega)$ that couple directly to 
the $J^{\charge}=2^{-}$ Bosonic modes. In this case 
the mass of the this Higgs mode is equal to the weak-coupling TDGL result, $M_{2,-} =\sqrt{12/5}\,\Delta$,
but now extended to all temperatures.
However, the weak-coupling result for the mass of the $J^{\charge}=2^{-}$ Higgs mode is renormalized by Fermionic 
interactions. Qualitatively this is expected given that external stress fluctuations couple directly to the 
$J^{\charge}=2^{-}$ pairing fluctuations. Excitation of a $J^{\charge}=2^{-}$ Higgs Boson {\sl polarizes} the $J=0$ 
Fermionic vacuum, inducing a Fermionic self-energy correction of the same symmetry that couples back to 
generate a mass correction to the $J^{\charge}=2^{-}$ Higgs modes.
The polarization correction to the Higgs mass is encoded in Eq. \ref{eq-Sigma+Scalar} in the limit $\vq=0$, 
which can be expressed as
\ber
\Sigma^{(+)}(\hp;\omega) &=& u^{(+)}(\hp;\omega) 
\nonumber\\
&+& \int\frac{d\Omega_{\hp'}}{4\pi}\,F^{s}(\hp,\hp')\,\Big\{
-
\lambda(\omega)\,\Sigma^{(+)}(\hp';\omega)\, 
\nonumber\\
&+& 
\nicefrac{1}{2}\bar\lambda(\omega)\,\omega\,\vec{\Delta}(\hp')\cdot\vec{d}^{(-)}(\hp';\omega)
\Big\}
\,,
\label{eq-Sigma+Scalar_q0}
\eer  
where $u^{(+)}(\hp;\omega)$ represents un-renormalized external forces coupling to excitations of \Heb{},
and we have expressed the dynamical self-energy in terms of the spin-symmetric particle-hole irreducible
interaction, $F^{s}(\hp,\hp')$ (c.f. Appendix \ref{sec-appendix_Dynamical_Equations}).
Projecting out the amplitudes with $J=0,2$ defined in Eq. \ref{eq-Sigma+Scalar_Jm} gives,
\ber
\left(1 + F^s_0\,\lambda(\omega)\right)\Sigma^{(+)}_{0,0}(\omega) 
&=& 
u_{0,0}(\omega) 
\label{eq-Sigma+Scalar_00}
\\
&+& 
F^s_0\,\lambda(\omega)\,\left(\frac{\omega}{2|\Delta|}\right)\,D^{(-)}_{0,0}(\omega)
\,,
\nonumber\\
\left(1 + \nicefrac{1}{5}F^s_2\,\lambda(\omega)\right)\Sigma^{(+)}_{2,m}(\omega) 
&=& 
u_{2,m}(\omega) 
\label{eq-Sigma+Scalar_2m}
\\
&+& 
\nicefrac{1}{5}
F^s_2
\,
\lambda(\omega)\,\left(\frac{\omega}{2|\Delta|}\right)
\,
D^{(-)}_{2,m}(\omega)
\,.
\nonumber
\eer
The key result shown in Eqs. \ref{eq-Sigma+Scalar_00}-\ref{eq-Sigma+Scalar_2m} is that excitation of pairing 
fluctuations, $D^{(-)}_{J,m}(\omega)$, polarizes the condensate and generates {\sl internal} stresses 
that are proportional to: (i) interactions in the particle-hole channel, $F^{s}_{2,0}$, (ii) the 
time-derivative of the Bosonic mode amplitudes, $\omega\,D^{(-)}_{J,m}(\omega)$, and (iii) the dynamical 
response of the condensate, $\lambda(\omega)$, even in the absence of bulk external forces, i.e. $u_{J,m}(\omega) = 0$.
In the case of the $J^{\charge}=0^{-}$ mode, combining Eq. \ref{eq-J=0-}, with $\Sigma^{(+)}_{0,0}(\omega)$ 
now given by Eq. \ref{eq-Sigma+Scalar_00} still yields the {\sl unrenormalized} dynamical equation for 
excitation of the Anderson-Bogoliubov phase mode,
\be
\omega^2\,D^{(-)}_{0,0} =  2\,|\Delta|\,\omega\,u^{(+)}_{0,0}
\,.
\ee
The interaction, $F^s_0$, drops out because the polarization induced by the $J^{\charge}=0^{-}$ Bosonic 
mode has the same rotational symmetry as the vacuum state.

However, in the case of the $J^{\charge}=2^{-}$ Higgs modes, combining 
Eq. \ref{eq-J=2-} with $\Sigma^{(+)}_{2,m}(\omega)$ given by Eq. \ref{eq-Sigma+Scalar_2m}, yields
\be
D^{(-)}_{2,m} 
=
\frac{\frac{4}{5}\,|\Delta|\,\omega\,u^{(+)}_{2,m}(\omega)}
{\Big[\omega^2 - \frac{12}{5}|\Delta|^2 + \lambda(\omega)\, \nicefrac{3}{25}\,F^{s}_{2}\, 
\left(\omega^2 - 4|\Delta|^2\right) \Big]}
\,,
\label{eq-J=2-Mode_vs_F2s}
\ee
which has a pole at $\omega=M_{2,-}$, the renormalized mass of the $J^{\charge}=2^{-}$ mode. Before discussing the
quantitative effect of the Landau interaction on the $J^{\charge}=2^{-}$ Higgs mass, we consider the effect of
interactions in the Cooper channel.

\vspace*{-3mm}
\subsection{F-wave interactions in the Cooper channel}
\vspace*{-3mm}
 
Theoretical models for fermionic interactions in the particle-particle (Cooper) channel based on 
exchange of long-lived ferromagnetic spin-fluctuations predict p-wave spin-triplet pairing with 
sub-dominant attraction in the f-wave Cooper channel, including a strong sub-dominant f-wave attractive
interaction at high pressures.\cite{lay74,fay75}
The masses of the $J^{\charge}=2^{-}$ modes are sensitive to fermionic interactions in the particle-particle 
channel, the most relevant being the f-wave, spin-triplet channel. Pairing fluctuations in the Cooper 
channel couple to the p-wave, spin-triplet modes with $J=2$ leading to re-normalization of the mass of 
the $J^{\charge}=2^{\pm}$ Higss modes. Note that f-wave pairing fluctuations do not couple to the $J=0,1$ 
Bosonic modes.

The generalization of Eqs. \ref{eq-J=2-} and \ref{eq-Sigma+Scalar_2m} to include the f-wave pairing channel in the 
dynamics of the $J^{\charge}=2^{-}$ modes is obtained from Eqs. \ref{eq-triplet-odd_parity-q0} and 
\ref{eq-Sigma+Scalar_q0} by retaining both p-wave and f-wave pairing amplitudes,
\be\label{eq-dminus_tensors}
d_{\alpha}^{(-)}(\hp) = \cD_{\alpha i}^{(-)}\,\hp_i + \cF_{\alpha; ijk}^{(-)}\,\hp_i\hp_j\hp_k
\,
\ee
where $\cD_{\alpha i}^{(-)}$ is a second-rank tensor under the residual symmetry group of the B-phase, 
$\point{SO(3)}{J}$, representing p-wave, spin-triplet fluctuations with odd charge conjugation parity 
(Eq. \ref{eq-p-wave-triplet_fluctuations}), and $\cF_{\alpha; ijk}^{(-)}$ is a fourth-rank tensor with 
f-wave orbital symmetry, and thus is completely symmetric and traceless in any pair of the orbital indices ($ijk$). 
Spin-triplet, f-wave pairing fluctuations couple only to the $J=2$ p-wave, triplet modes. Thus, for 
pure $J=2$, $S=1$, $\ell=3$ fluctuations,
\ber
\cF_{\alpha;ijk}^{(-)} 
=
\nicefrac{5}{9}
\left\{
\right.
&& 
\hspace*{-4mm}
\left(
\delta_{\alpha i}\cF_{jk}^{(-)}
+
\delta_{\alpha j}\cF_{ik}^{(-)}
+
\delta_{\alpha k}\cF_{ij}^{(-)}
\right) 
\nonumber
\\
-
\nicefrac{2}{5}
&&
\left.
\hspace*{-4mm}
\left(
\cF_{\alpha i}^{(-)}\delta_{jk} 
+
\cF_{\alpha j}^{(-)}\delta_{ik} 
+
\cF_{\alpha k}^{(-)}\delta_{ij} 
\right) 
\right\}
\,,
\eer
where by contraction  
\be
\cF^{(-)}_{jk}\equiv \nicefrac{3}{7}\,\cF^{(-)}_{\alpha;\alpha jk}
\,,
\ee
is a rank two, traceless and symmetric $J=2$ tensor. In particular, we can expand $\cF^{(-)}_{ij}$ in the $J=2$ base tensors,
\be
\cF^{(-)}_{ij} =  \sum_{m=-2}^{+2}\,F^{(-)}_{2,m}\,t^{(2,m)}_{ij}
\,.
\ee
The $J=2^{-}$ gap distortion is determined by both the p- and f-wave $J=2$ tensors,
\be
\vec\Delta(\hp)\cdot\vec{d}^{(-)}(\hp;\omega) = \Delta\,\left(\cD^{(-)}_{ij} + \cF^{(-)}_{ij}\right)\hp_i\hp_j
\,,
\ee
and thus the $J=2^{-}$ component of the Fermionic self-energy induced by the $J^{\charge}=2^{-}$ Higgs 
modes (c.f. Eq. \ref{eq-Sigma+Scalar_2m}) becomes,
\ber
\left(1 + \nicefrac{1}{5}F^s_2\,\lambda(\omega)\right)\Sigma^{(+)}_{2,m}(\omega) 
&=& 
u_{2,m}(\omega) + \nicefrac{1}{5} F^s_2\,\lambda(\omega)
\label{eq-Sigma+Scalar_2m-p+f}
\\
&\times& 
\left(\frac{\omega}{2|\Delta|}\right)
\left[
D^{(-)}_{2,m}(\omega)
+
F^{(-)}_{2,m}(\omega)
\right]
\,.
\nonumber
\eer

The $J^{\charge}=2^{-}$ amplitudes satisfy coupled time-dependent gap equations obtained by projecting out
the p- and f-wave components of Eq. \ref{eq-triplet-odd_parity-q0} where $v_{\ell}$ are the pairing interactions in orbital 
angular momentum channel, $\ell = 1, 3, \ldots$. The p-wave interaction is the dominant attractive channel.
The relevant measure of the strength of the sub-dominant f-wave pairing interaction is
\be
x_3^{-1} \equiv \left(\frac{1}{v_1} - \frac{1}{v_3}\right)^{-1} = \ln\left(T_{c_3}/T_c\right)^{-1}
\,,
\ee
where $x_3^{-1} < 0$ ($x_3^{-1} > 0$) for attractive (repulsive) f-wave pairing. The latter equality, valid for attractive 
f-wave pairing, is obtained from Eq. \ref{eq-Linearized_Gap-Equation} for the eigenvalue spectrum of the linearized gap 
equation, with $T_c$ the p-wave transition temperature and $T_{c_3}$ the f-wave {\sl instability 
temperature} for sub-dominant f-wave pairing.

Projecting out the $\ell =1$, $J=2$ component of Eq. \ref{eq-triplet-odd_parity-q0}, which generalizes 
Eq. \ref{eq-J=2-}, leads to
\be\label{eq-J=2-p}
\Big[\omega^2 - \nicefrac{12}{5}|\Delta|^2\Big]\,D^{(-)}_{2,m} + \nicefrac{8}{5}|\Delta|^2\,F^{(-)}_{2,m} 
=  
\nicefrac{4}{5}\,|\Delta|\,\omega\,\Sigma^{(+)}_{2,m}
\,.
\ee
Projecting out the $\ell=3$ amplitudes from Eq. \ref{eq-triplet-odd_parity-q0} gives
\ber
\label{eq-J=2-f1}
\Big[
x_3 + \nicefrac{1}{4}\bar\lambda(\omega)(\omega^2 - 4|\Delta|^2)
\Big]
\,
d^{(3,-)}_{\alpha}(\hp;\omega) 
\hspace*{2cm}
&&
\\
+ 
7\int\frac{d\Omega_{\hp'}}{4\pi}\,P_3(\hp\cdot\hp')\,
\left\{
\bar\lambda(\omega)\,
\Delta_{\alpha}(\hp')\,
\vec{\Delta}(\hp')\cdot\vec{d}^{(-)}(\hp';\omega)
\right.
&&
\nonumber
\\
\left.
=
\nicefrac{1}{2}\bar\lambda(\omega)\,\omega\,\Delta_{\alpha}(\hp')\,\Sigma^{+}(\hp';\omega)
\right\}
&&
\,.
\nonumber
\eer

The $J=2$ components of Eq. \ref{eq-J=2-f1} are obtained by contracting with $\hp_{\alpha}$ to obtain an equation for 
$F^{(-)}(\hp)\equiv \cF^{(-)}_{ij}\hp_i\hp_j$, then evaluating the angular average using the addition theorem for the 
Legendre polynomials, 
$(\hp\cdot\hp')\,P_3(\hp\cdot\hp') = \nicefrac{1}{7}\left\{ 4\,P_4(\hp\cdot\hp') + 3\,P_2(\hp\cdot\hp')\right\}$, 
to obtain
\be
\label{eq-J=2-f}
\hspace*{-3mm}
\Big[
\bar{x}_3 + \nicefrac{1}{4}(\omega^2 - \nicefrac{8}{5}|\Delta|^2)
\Big]
F_{2,m}^{(-)}
+ 
\nicefrac{3}{5}|\Delta|^2\,D^{(-)}_{2,m}
=
\nicefrac{3}{10}\,|\Delta|\,\omega\,\Sigma_{2,m}^{+}
\,,
\ee
\vspace*{1mm}

\noindent where $\bar{x}_3\equiv x_3/\bar\lambda(\omega)$. Eliminating the Fermionic self-energy between Eqs. \ref{eq-J=2-p} 
and \ref{eq-J=2-f} gives the sub-dominant f-wave, $J^{\charge}=2^{-}$ amplitude in terms of the dominant 
p-wave, $J^{\charge}=2^{-}$,
\be
\label{eq-J=2-p+f}
\Big[
\bar{x}_3 + \nicefrac{1}{4}(\omega^2 - 4|\Delta|^2)
\Big]
F_{2,m}^{(-)}
=
\nicefrac{3}{8}(\omega^2 - 4|\Delta|^2)\,D^{(-)}_{2,m}
\,.
\ee

The {\it total} $J^{\charge}=2^{-}$ Higgs amplitude - the sum of the p- and f-wave amplitudes, 
$H^{(-)}_{2,m}(\omega) \equiv D^{(-)}_{2,m} + F^{(-)}_{2,m}$ -  that polarizes the Fermionic vacuum
(Eq. \ref{eq-Sigma+Scalar_2m})
is governed by the dynamical equation obtained by combining Eqs. \ref{eq-J=2-p} and \ref{eq-J=2-p+f}. 
This gives the retarded propagator for the $J^{\charge}=2^{-}$ Higgs mode,

\begin{widetext}
\be\label{eq-J=2-Mode_vs_F2s+x3i}
H^{(-)}_{2,m} 
=
\frac{
\frac{4}{5}\,|\Delta|\,\omega\,u^{(+)}_{2,m}(\omega)\,
		  \left[1+\nicefrac{5}{8}\,x_3^{-1}\,(\omega^2-4|\Delta|^2)\bar\lambda(\omega)\right]
     }
{
\Big[\omega^2 - \frac{12}{5}|\Delta|^2 + \lambda(\omega)\,(\omega^2 - 4|\Delta|^2)\,
	\left(
		\nicefrac{3}{25}\,F^{s}_{2}\, + ({\omega}/{2|\Delta|})^2\,x_3^{-1}
	\right)
\Big]
}
\,.
\ee
\end{widetext}

%---------------- Figure: Mass corrections J=2 -----------------------
\begin{figure}[t]
\begin{center}
\includegraphics[width=\linewidth]{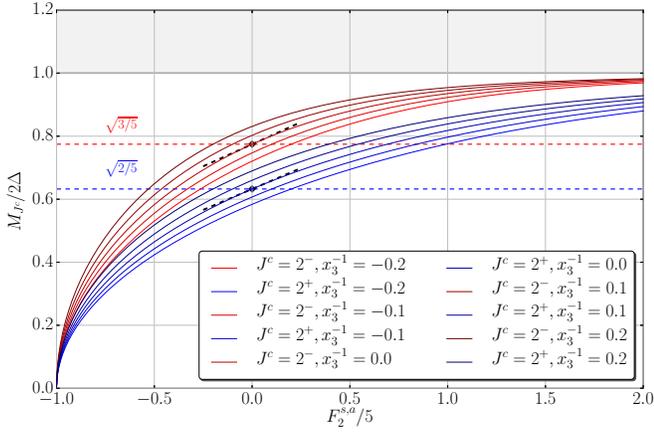}
\caption{Masses of the $J^{\charge}=2^{\pm}$ Higgs modes vs. $\ell =2$ particle-hole ($F_2^{s,a}$) and 
         f-wave pairing ($x_3^{-1}$) interactions at $T=0$. The perturbative results 
         (Eqs. \ref{eq-Mass_J=2-_perturbative-F2s+x3i} and \ref{eq-Mass_J=2+_perturbative-F2a+x3i}) 
	 for $x_3^{-1}=0$ are shown as the dashed black lines.
	}	
\label{fig-Modes_vs_F2sa+x3i}
\end{center}
\end{figure}
%---------------------------------------------------------------------

The renormalized $J^{\charge}=2^{-}$ Higgs mass is obtained from the pole of the  
propagator in Eq. \ref{eq-J=2-Mode_vs_F2s+x3i}.
In the limit $T\rightarrow T_c^{-}$ the Tsuneto function scales as 
$\lambda(\omega=M_{2,-}) \propto \Delta(T)/T_c\rightarrow 0$. Thus, the $J^{\charge}=2^{-}$ Higgs mass 
scales to the the weak-coupling TDGL result at $T_c$,
\be\label{eq-Mass_J=2-_limit-Tc}
M_{2^-} \approx \sqrt{\frac{12}{5}}\Delta(T)\,
\Bigg[1 + \frac{\pi}{10}\sqrt{\frac{5}{2}} \,\frac{\Delta(T)}{T_c}\,(F^{s}_{2}/5 + x_3^{-1})\Bigg]
\,,
\ee
However, the leading order correction to the mass, $\propto \Delta(T)/T_c\sim(1-T/T_c)^{\nicefrac{1}{2}}$, onsets 
rapidly below $T_c$. Thus, mass renormalzation becomes significant, of order $F_2^2$ or $x_3^{-1}$, 
for $T\rightarrow 0$.
For weak interactions in both the Landau and Cooper channels, $|F_2^{s}| \ll 1$ and $|x_3^{-1}|\ll 1$, at $T=0$ the 
renormalized mass obtained from the pole of the $J^{\charge}=2^{-}$ propagator in 
Eq. \ref{eq-J=2-Mode_vs_F2s+x3i} is
\be\label{eq-Mass_J=2-_perturbative-F2s+x3i}
M_{2^-} \approx \sqrt{\frac{12}{5}}\Delta\,\Bigg[1 + a\,(F^{s}_{2}/5 + x_3^{-1})\Bigg]
\,,
\ee
where $a=\nicefrac{1}{\sqrt{6}}\arcsin\left(\sqrt{\nicefrac{3}{5}}\right)\approx 0.362$. The Landau channel 
interaction $F_2^s$ obtained from measurements of the zero sound velocity ranges from 
$F_2^s\approx 0.5$ at $P=0\,\mbox{bar}$ to 
$F_2^s\approx 1.0$ at $P=34\,\mbox{bar}$, although earlier measurements reported $F_2^s \approx -0.5$ 
at $p=0\,\mbox{bar}$.\cite{har00}

The f-wave interaction in the Cooper has been determined from measurements of the mass of 
the $J^{\charge}=2^{-}$, $m=0$ Higgs mode based on resonant absorption of longitudinal zero sound. 
These experiments yield results ranging from $x_3^{-1}\approx 0.0$ at $p=0\,\mbox{bar}$ to 
$x_3^{-1}\approx -0.5$ at $p=14\,\mbox{bar}$ (c.f. Fig. 50 in Ref. \onlinecite{hal90}).
Determinations of the mass of the $J^{\charge}=2^{-}$, $m=\pm 1$ Higgs modes based on transverse sound propagation 
and acoustic Faraday rotation by Lee et al.\cite{lee99,sau00a}, as well as more recent measurements by 
Collett et al.\cite{col13} yield attractive f-wave interactions of similar magnitude. 
The f-wave interaction in the Cooper channel also contributes to the nonlinear nuclear magnetic susceptibility 
for the B-phase.\cite{fis86} Analysis of magnetic susceptibility measurements of Hoyt et al.\cite{hoy81} yields a 
stronger, but sub-dominant, attractive f-wave interaction with $x_3^{-1}\simeq -1.75$ ($T_{c_3}/T_c\simeq 0.56$) 
at low pressure.\cite{fis88}

Figure \ref{fig-Modes_vs_F2sa+x3i} shows the mass of the $J=2^{\charge}=2^{-}$ Higgs mode as a function 
of $F^{s}_2$ for various values of the f-wave pairing interaction, $x_3^{-1}$, obtained from numerical 
solution for the pole of the propagator, $H^{(-)}_{2,m}$, in Eq. \ref{eq-J=2-Mode_vs_F2s+x3i}.
Note that `repulsive' interactions in either channel ($F_2^{s,a} > 0$ or $x_3^{-1}>0$) push the mass above the 
weak-coupling result towards the mass of un-bound Fermion pairs, while 'attractive interactions' soften the mode. 
In particular, $M_{2,-}\rightarrow 0$ for $F_2^{s}/5\rightarrow -1$, signaling a dynamical instability 
of the ground-state. The soft mode is the dynamical signature of the Pomeranchuk instability of the underlying 
Fermionic vacuum.\cite{baym91}

\vspace*{-3mm}
\subsection{Nambu-Goldstone and Higgs Modes with $\charge=+1$}\label{sec-J+Modes}
\vspace*{-3mm}

In the case of the Bosonic modes with parity $\charge=+1$ the Fermion self-energy
that couples to these modes is expressed in terms of the momentum-dependent exchange field,
$\vec\Sigma^{(+)}(\hp;\omega)$.
Equation \ref{eq-triplet-even_parity-q0} decouples into the dynamical equations for Bosonic mode 
amplitudes with total angular momentum $J$, with orbital angular momentum $\ell=1$, $D^{(+)}_{J,m}$, 
and $\ell=3$, $F^{(+)}_{J,m}$.
The self-energy fluctuations originating from the exchange contribution to the quasiparticle 
interaction are even 
under $\hp\rightarrow-\hp$; thus only fluctuations with even $J$ couple to the Bosonic modes for 
$\charge =+1$. To obtain the dynamical equations for the $J^{+}$ modes it is convenient to introduce
\be
\vec{G}^{(+)}(\hp;\omega) 
= 
\vec\Delta(\hp)\times\vec\Sigma^{(+)}(\hat{p};\omega)/|\vec\Delta(\hp)|
\,,
\label{eq_G+}
\ee
For the $J=0^{+}$ ground state $\vec{G}^{(+)}(\hp;\omega)=\hp\times\vec{\Sigma}^{(+)}$
is a vector under spin rotations, \emph{odd} under $\hp\rightarrow -\hp$ and enters 
Eq. \ref{eq-triplet-even_parity-q0} acts as an effective source field for Cooper pair 
fluctuations with $\charge=+1$.

It is sufficient to retain only the $\ell=0$ and $\ell=2$ contributions to the particle-hole
exchange interaction, $F^{a}(\hp,\hp') = F^{a}_0 + F^{a}_2\,P_2(\hp\cdot\hp)$, in which case
we can express the vector components of the quasi-particle exchange field in terms 
$\ell=0$ and $\ell=2$ spherical tensors,
\be
\Sigma_{\gamma}^{(+)}(\hp;\omega) 
= 
\Sigma_{\gamma}^{(0)} + \Sigma^{(2)}_{\gamma:ij}\,\hp_{i}\hp_{j}
\,,
\ee
where $\Sigma^{(2)}_{\gamma:ij}$ is traceless and symmetric in the indices, $ij$.
The vector function, $\vec{G}^{(+)}(\hp;\omega)$, by construction contains only p-wave and f-wave 
orbital components, $\vec{G}^{(+)}(\hp;\omega)=\vec{G}^{(1)}(\hp;\omega)+\vec{G}^{(3)}(\hp;\omega)$, with 
$G^{(\ell)}_{\gamma}(\hp)=\langle(2\ell+1)\,P_{\ell}(\hp\cdot\hp')\,G^{(+)}_{\gamma}(\hp')\rangle_{\hp'}$,
where $\langle\ldots\rangle_{\hp}\equiv\int d\Omega_{\hp}/4\pi (\ldots)$.
Equivalently, the p-wave contribution is defined by a second-rank tensor under joint 
spin- and orbital rotations,
\ber
G^{(1)}_{\gamma i}
&=&
\langle 3\,\hp_i\,G^{(+)}_{\gamma}(\hp)\rangle_{\hp}
=
\varepsilon_{\alpha i \gamma}\Sigma_{\gamma}^{(0)} 
+ 
\frac{2}{5}\varepsilon_{\alpha j\gamma}\Sigma^{(2)}_{\gamma:ji}
\\
&=&
G^{(1,0)}_{\gamma i} + G^{(1,1)}_{\gamma i} + G^{(1,2)}_{\gamma i}
\,,
\label{eq_G_p-wave_J}
\eer
where the second equation is the reduction in terms of $J=0,1,2$ tensors.
The $J=0$ component is defined by the trace, which is easily seen to vanish, i.e. 
$G^{(1,0)}_{\alpha i} \equiv 0$. The $J=1$ components can be expressed in terms of an axial vector, 
\be\label{eq_G_11}
G^{(1,1)}_{\alpha i} = \varepsilon_{\alpha i \nu}\,G^{(1,1)}_{\nu}\,\,
\mbox{with}\,\,G^{(1,1)}_{\nu}=\Sigma^{(0)}_{\nu} - \nicefrac{1}{5}\,\Sigma^{(2)}_{\gamma:\gamma\nu}
\,.
\ee 
Finally, the $J=2$ components are determined by the traceless, symmetric tensor
\be\label{eq_G_12}
G^{(1,2)}_{\alpha i} = 
\nicefrac{1}{5}
\left(
\varepsilon_{\alpha j\gamma}\,\Sigma^{(2)}_{\gamma:j i}
+
\varepsilon_{ij\gamma}\,\Sigma^{(2)}_{\gamma:j \alpha}
\right)
\,,
\ee
which can be expanded in the basis of $J=2$ tensors, 
\be
G^{(1,2)}_{\alpha i} = \sum_{m=-2}^{+2}\,G_{2,m}\,t^{(2,m)}_{\alpha i}
\,.
\ee 
These contributions to the exchange field couple to the Bosonic mode amplitudes with quantum 
numbers, $J,m$ and $\charge=+1$, represented by second- and fourth-rank tensors that are the 
$\charge=+1$ complements of those in Eq. \ref{eq-dminus_tensors},
\be\label{eq-dplus_tensors}
d_{\alpha}^{(+)}(\hp) = \cD_{\alpha i}^{(+)}\,\hp_i + \cF_{\alpha; ijk}^{(+)}\,\hp_i\hp_j\hp_k
\,,
\ee
where the spin-triplet, p-wave order parameter fluctuations are expanded in the basis 
of tensors with $J=0,1,2$,
\be
\cD_{\alpha i}^{(+)} = \sum_{J=0,1,2}\sum_{m=-J}^{J}\,D_{J,m}^{(+)}\,t^{(J,m)}_{\alpha i}
\,,
\ee
and similarly for spin-triplet, f-wave fluctuations with $J=2^{+}$,
$\cF_{\alpha i}^{(+)} = \sum_{m=-2}^{+2}\,F_{2,m}^{(+)}\,t^{(2,m)}_{\alpha i}$
where $\cF_{\alpha i}^{(+)} = \nicefrac{3}{7}\cF_{\gamma:\gamma\alpha i}$

The equation governing the $J^{\charge}=0^{+}$ mode is
\be\label{eq-J=0+}
\left( \omega^2\, - 4|\Delta|^2 \right) D^{(+)}_{0,0} =  0
\,.
\ee
This is the dynamical equation for the Higgs mode with the \emph{exact} quantum numbers of the 
B-phase vacuum state. As a result there is no coupling to the $J^{\charge}=0^{+}$ mode via 
acoustic or magnetic fluctuations.\footnote{However, the process of two-phonon absorption and 
excitation of the $J^{\charge}=0^{+}$ is not forbidden.}

The $J^{\charge}=1^{+}$ modes are Nambu-Goldstone modes associated with broken \emph{relative} 
spin-orbit rotation symmetry. 
It is convenient to express these mode amplitudes in the Cartesian representation,
$D^{(+,1)}_{\alpha}=\nicefrac{1}{2}\varepsilon_{\alpha\beta\gamma}\,\cD^{(+)}_{\beta\gamma}$. 
Projecting out these amplitudes from Eq. \ref{eq-triplet-even_parity-q0} yields,
\be\label{eq_J=1+}
i\omega\,D^{(+,1)}_{\alpha} 
=
2\Delta\,\frac{1+\nicefrac{1}{15}\lambda(\omega)\,F_2^a}
                        {1-\nicefrac{2}{45}\lambda(\omega)^2\,F_0^a\,F_2^a}\,
\left(-\nicefrac{\gamma\hbar}{2}\,H_{\alpha}(\omega)\right)
\,,
\ee
where $H_{\alpha}(\omega)$ is the Fourier component of the time-dependent external magnetic field 
and $\gamma$ is the gyromagnetic ratio of \He.
Exchange interactions renormalize the coupling of the $J^{\charge}=1^{+}$ modes to an external field, but  
the massless NG mode is protected by the continuous degeneracy of the BW ground state with respect to 
\emph{relative} spin-orbit rotations.  
At finite wavelength these excitations correspond spin waves mediated by $J^{\charge}=1^{+}$ NG modes of the 
Cooper pairs with dispersion given by $\omega=c_{m}\,q$, where $c_m$ are the spin-wave velocities in \Heb.
See Sec. \ref{sec-Light_Higgs} discussion of weak symmetry-breaking perturbations on the $J^{\charge}=1^{+}$
modes.

The $J^{\charge}=2^{+}$ excitations obey the dynamical equations,
\be\label{eq_p-wave_J=2+}
\Big[\omega^2 - \frac{8}{5}|\Delta|^2\Big]\,D^{(+)}_{2,m} 
=  
\frac{8}{5}|\Delta|^2\,F^{(+)}_{2,m} 
-i\omega(2\Delta)\,G_{2,m}
\,.
\ee
In the absence of Fermion interactions in the particle-particle channel the f-wave amplitude 
vanishes, $F^{(+)}_{2,m}\equiv 0$. And, if we also ignore Fermionic interactions in the particle-hole channel, 
then $G_{2,m}(\omega)$ represents an {\sl external} field that couples to directly to the 
$J^{\charge}=2^{+}$ modes. In this case the mass of this Higgs mode is equal to 
the weak-coupling result, $M_{2,+} =\sqrt{8/5}\,\Delta$.
However, $M_{2,+}$ is renormalized by Fermionic interactions in both the particle-particle and particle-hole 
channels. Just as in the case for the $J^{\charge}=2^{-}$ modes excitation of a $J^{\charge}=2^{+}$ Higgs 
Boson {\sl polarizes} the $J=0^{+}$ Fermionic vacuum and introduces a Fermionic self-energy correction with the 
same symmetry that couples back to generate a mass correction to the $J^{\charge}=2^{+}$ Higgs modes.

In addition, pairing interactions in the spin-triplet, f-wave channel lead to dynamical excitations of the 
B-phase vacuum with spin $J^{\charge}=2^{+},m$, i.e. $F^{(+)}_{2,m}$, which mixes with the spin-triplet, p-wave 
modes of the same symmetry. We obtain the dynamical equation for the $F^{(+)}_{2,m}$ amplitudes by 
projecting out the f-wave orbital components of Eq. \ref{eq-triplet-even_parity-q0} to obtain,
\be\label{eq_f-wave_J=2+}
\Big[4\bar{x}_3 + (\omega^2 - \frac{12}{5}|\Delta|^2)\Big]\,F^{(+)}_{2,m} 
+
\frac{12}{5}|\Delta|^2\,D^{(+)}_{2,m} 
=  
+i\omega(2\Delta)\,G_{2,m}
\,,
\ee
where $\bar{x}_3\equiv x_3/\bar{\lambda}(\omega)$. Note that we have used the identity, 
$\hp\cdot\vec{G}(\hp)= \hp\cdot\vec{G}^{(1)}(\hp)+ \hp\cdot\vec{G}^{(3)}(\hp)\equiv 0$ 
to express the source term in Eq. \ref{eq_f-wave_J=2+} in terms of the p-wave, $J=2$ component 
of $\vec{G}(\hp)$, i.e. $G^{(3,2)}_{\gamma i} = - G^{(1,2)}_{\gamma i}$. 
Eliminating $G_{2,m}$ from Eqs. \ref{eq_p-wave_J=2+} and \ref{eq_f-wave_J=2+} gives the f-wave,
$J^{\charge}=2^+$ amplitude in terms of the corresponding dominant p-wave amplitude,
\be
\label{eq_F-D_J=2+}
\Big[4\bar{x}_3 + (\omega^2 - 4|\Delta|^2)\Big]\,F^{(+)}_{2,m}
=
-(\omega^2 - 4|\Delta|^2)\,D^{(+)}_{2,m}
\,.
\ee

The polarization corrections to the $J^{\charge}=2^+$ Higgs mass are obtained from Eqs. 
\ref{eq_p-wave_J=2+}, \ref{eq_F-D_J=2+} and \ref{eq_Sigma-Vector-Plus} in the limit $\vq=0$, 
which can be expressed as
\ber
\vec{\Sigma}^{(+)}(\hp;\omega) &=& \vec{h}^{(+)}(\hp;\omega) 
+
\int\frac{d\Omega_{\hp'}}{4\pi}\,F^{a}(\hp,\hp')\,\Big\{
\nonumber\\
&-& 
\lambda(\omega)\,
\left(
\vec{\Sigma}^{(+)}(\hp';\omega)\,
-
\hp'(\hp'\cdot\vec{\Sigma}^{(+)}(\hp';\omega))
\right)
\nonumber\\
&-& 
\left(\frac{\omega}{2\Delta}\right)\lambda(\omega)\,\hp'\times\vec{d}^{(+)}(\hp';\omega)
\Big\}
\,,
\label{eq-Sigma+Vector_q0}
\eer  
where $\vec{h}^{(+)}(\hp;\omega)$ represents the external field coupling to Fermionic excitations 
via the magnetic moment of the \He\ nucleus, and $F^{a}(\hp,\hp')$ represents the spin-dependent 
exchange interaction in \He{} (c.f. Eq. \ref{eq_Gamma-ph}, the paragraph preceding 
Eq. \ref{eq-ph_scattering-amplitude} and Eq. \ref{eq_forward-scattering_Landau}. 
Note that Eq. \ref{eq_forward-scattering_Landau} has been inverted and used to 
express $\vec{\Sigma}^{(+)}(\hp;\omega)$ in
terms of the Landau interaction, $F^{a}(\hp,\hp')$.
For $\charge=+1$ Bosonic excitations the coupling of the Fermionic self-energy fluctuations 
is determined by the p-wave, $J=0,1,2$ components of $\vec{G}^{(+)}(\hp;\omega)$
in Eq. \ref{eq_G_p-wave_J}. Fluctuations of $G^{(1)}_{\gamma i}$ with $J=0$ vanish by symmetry as
$\vec{G}^{(+)}(\hp;\omega)$ is purely transverse with respect to $\hp$. Fluctuations with $J=1$ are
defined by the $\ell=0,2$ orbital components of $\vec{\Sigma}^{(+)}(\hp;\omega)$ in 
Eq. \ref{eq_G_11}, while the $J=2$ components are defined by 
Eq. \ref{eq_G_12}. 

The dynamical equation for 
$G^{(1,1)}_{\alpha i}= \varepsilon_{\alpha i \gamma}
\left(\Sigma^{(0)}_{\gamma}-\nicefrac{1}{5}\Sigma^{(2)}_{\nu:\nu\gamma}\right)$ 
is constructed from the equations for the $\ell=0$ and $\ell =2$ exchange fields,
\ber
\left(1+\nicefrac{2}{3}\lambda F^{a}_0\right)\,\Sigma^{(0)}_{\gamma} 
&=& 
\hspace*{-2mm}
h_{\gamma} 
-
i\nicefrac{2}{3}\left(\frac{\omega}{2\Delta}\right)\lambda\,F^a_0\,
\nicefrac{1}{2}\varepsilon_{\alpha\beta\gamma}\,\cD^{(1,1)}_{\alpha\beta}
\ns,
\quad
\\
\hspace*{-12mm}
\left(1+\nicefrac{1}{15}\lambda F^{a}_2\right)\,\Sigma^{(2)}_{\nu:\nu\gamma} 
&=& 
\hspace*{-2mm}
i\nicefrac{1}{3}\left(\frac{\omega}{2\Delta}\right)\lambda\,F^a_2\,
\nicefrac{1}{2}\varepsilon_{\alpha\beta\gamma}\,\cD^{(1,1)}_{\alpha\beta}
\,,
\eer
which shows that $G^{(1,1)}_{\alpha i}$ couples \emph{only} to the $\ell=1$, 
$J^{\charge}=1^{+}$ Bosonic modes, thus leading to Eq. \ref{eq_J=1+} for these NG modes. 

The Fermionic self-energy that couples to the $J^{\charge}=2^{+}$ Bosonic modes is determined
by the $\ell=2$ components of the exchange field defined by $G^{(1,2)}_{\alpha i}$ in 
Eq. \ref{eq_G_12}. 
The equation of motion for the $(2,m)$ components are then 
\be\label{eq_G_12m}
\left(1+\nicefrac{1}{5}\lambda F^{a}_2\right)\,G_{2,m}
=
h^{(1,2)}_{2,m} - i\left(\nicefrac{\omega}{2\Delta}\right)\nicefrac{3}{25}\lambda\,F^a_2\,
\left\{2F^{(+)}_{2,m}-D^{(+)}_{2,m}\right\}
\,,
\ee
where $h^{(1,2)}_{2,m}$ are the components of a generalized, momentum-dependent, external magnetic 
field that couples to Fermionic and Bosonic excitations with $J=2$ via the nuclear spin. 
Combining Eqs. \ref{eq_p-wave_J=2+}, \ref{eq_F-D_J=2+} and \ref{eq_G_12m} we obtain the response 
function for the $J^{\charge}=2^{+}$ Higgs amplitude, $H^{(+)}_{2,m}=D^{(+)}_{2,m} + F^{(+)}_{2,m}$,

\begin{widetext}
\be\label{eq-J=2+Mode_vs_F2a+x3i}
H^{(+)}_{2,m}= 
\frac{-i\omega(2|\Delta|)\,h^{(1,2)}_{2,m}(\omega)}
{(\omega^2-\nicefrac{8}{5}|\Delta|^2)
 +\lambda(\omega)(\omega^2-4|\Delta|^2)
  \left(\nicefrac{2}{25}\,F^{a}_{2} + \left(\omega/2|\Delta|\right)^2\,x_3^{-1}\right)
}
\,.
\ee
\end{widetext}

The renormalized mass of the $J^{\charge}=2^{+}$ Higgs mode is obtained from the pole of the  
propagator in Eq.  \ref{eq-J=2+Mode_vs_F2a+x3i};
$M_{2^+}$ scales to the the weak-coupling TDGL result for $T\rightarrow T_c^{-}$, 
\be\label{eq-Mass_J=2+_limit-Tc}
M_{2^+} \approx \sqrt{\frac{8}{5}}\Delta(T)\,
\Bigg[1 + \frac{\pi}{4}\sqrt{\frac{3}{5}}\,\frac{\Delta(T)}{T_c}\,(F^{a}_{2}/5 + x_3^{-1})\Bigg]
\,,
\ee
with the leading-order correction developing rapidly below $T_c$. For weak interactions, 
$|F_2^{s}| \ll 1$ and $|x_3^{-1}|\ll 1$, the vacuum polarization correction at $T=0$ can also be 
calculated perturbatively,
\be\label{eq-Mass_J=2+_perturbative-F2a+x3i}
M_{2^+} \approx \sqrt{\frac{8}{5}}\Delta\,\Bigg[1 + b\,(F^{a}_{2}/5 + x_3^{-1})\Bigg]
\,,
\ee
where $b=\nicefrac{3}{2\sqrt{6}}\,\arcsin\left(\sqrt{\nicefrac{2}{5}}\right)\approx 0.419$. 
Note that the $\ell=2$ exchange interaction, $F_2^a$, is reported by Halperin and Varoquaux \cite{hal90} 
to vary between $F_2^a\approx -0.88$ at $P=0\,\mbox{bar}$ and $F_2^a\approx -0.01$ at $P=32\,\mbox{bar}$.
Figure \ref{fig-Modes_vs_F2sa+x3i} shows the mass of the $J^{\charge}=2^{+}$ Higgs mode as a function of 
the the $\ell=2$ exchange interaction, $F_2^{a}$, for various values of the f-wave interaction, $x^{-1}_3$ 
obtained from numerical solution for the pole of the propagator, $H^{(+)}_{2,m}$, in 
Eq. \ref{eq-J=2+Mode_vs_F2a+x3i}. 
Repulsive interactions push the mass above the weak-coupling result. Attractive f-wave and exchange 
interactions reduce the mass; the f-wave interaction is less effective for strong 
ferromagnetic exchange, $F_2^a/5\rightarrow -1$, for which $M_{2^+}\rightarrow 0^+$, as is clear from the 
equation for $M_{2^+}$ defined by the pole of Eq. \ref{eq-J=2+Mode_vs_F2a+x3i}. In this limit the soft 
mode is dominated by the Pomeranchuk instability of the underlying Fermionic vacuum. Nevertheless, for 
fixed $F_2^a/5 > -1$ $M_{2^+}\rightarrow 0^+$ as $T_{c_3}\rightarrow T_{c}$ ($x_3^{-1}\rightarrow -\infty$).

%-----------------------------
\begin{figure}[t]
\begin{center}
\includegraphics[width=\linewidth]{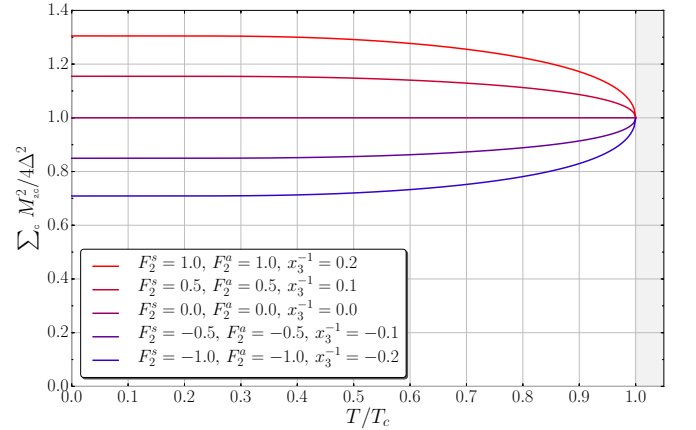}
\caption{Deviation of the Nambu Sum from polarization corrections to the the $J=2$ Higgs modes of \Heb\ 
for a range of interactions in both the Landau and Cooper channels.
}
\label{fig-NSR_vs_FsaX3i}
\end{center}
\end{figure}
%-----------------------------
 
The charge conjugation parity of the Bosonic modes with the same orbital, spin and total angular momentum
quantum number is reflected dramatically in the polarization corrections to the masses of the Higgs
modes. The $J^{\charge}=2^{-}$ modes couple to a quadrupolar excitation of the Fermionic vacuum, leading 
to a mass shift from the interaction $F_2^{s}$ in the spin-symmetric particle-hole channel,
which is generally repulsive except possibly near $p=0\,\mbox{bar}$.\cite{hal90} By contrast excitation of 
the $J^{\charge}=2^{+}$ modes is coupled to a quadrupolar spin-polarization, and thus has a
polarization correction to its mass from the interaction $F_2^{a}$ in the anti-symmetric (exchange) 
particle-hole channel; this interaction is expected to be attractive at all pressures. In addition,
both $J^{\charge}=2^{\pm}$ Higgs modes couple to f-wave pairing fluctuations with the same $J$
and parity $\charge$. In this case the asymmetry in the mass shifts for $J^{\charge}=2^{\pm}$
originates from $(\omega/2|\Delta|)^2\,x_3^{-1}$. Thus, the aysmmetry in the weak-coupling mass
spectrum, i.e. $\sqrt{12/5}\Delta$ vs. $\sqrt{8/5}\Delta$, leads to additional asymmetry in
the polarization corrections from the f-wave interactions in the Cooper channel. These 
trends are shown explicitly by the perturbative results in Eqs. \ref{eq-Mass_J=2-_perturbative-F2s+x3i} 
and \ref{eq-Mass_J=2+_perturbative-F2a+x3i}. Figure \ref{fig-NSR_vs_FsaX3i} summarizes the magnitude of the 
corrections to the NSR for a range of interactions in the Landau and Cooper channels. The violation
of the NSR onsets rapidly below $T_c$, with deviations of order $20-30\%$ for the Fermionic interactions 
characteristic of normal \He.

Excitation of the $J^{\charge}=2^{+},m$ modes typically occurs through weakly coupled channels at 
finite wavelength, $q\ne 0$, as coupling via an external field with symmetry $h_{2,m}^{(1,2)}$ is not
easily realized.
Koch and W\"olfle showed that the weak violation of particle-hole symmetry by the normal-state 
Fermionic vacuum lifts a selection rule that otherwise prohibits the coupling of the 
$J^{\charge}=2^{+}$ Higgs modes to density and mass current fluctuations.\cite{koc81} 
Thus, the $J^{\charge}=2^{+}, m$ modes can be excited by density and mass current channels, albeit 
with a coupling that is reduced by the factor, $\zeta\approx k_{\text{B}}T_c/E_f\ll 1$, 
the measure of the asymmetry of the spectrum of particle and hole excitations of the normal 
Fermionic vacuum at $\varepsilon\approx k_{\text{B}}T_c$.\cite{mck90}
This coupling leads to resonant excitation of the $J^{\charge}=2^{+}$ Higgs mode by absorption of 
zero-sound phonons. Indeed ultra-sound absorption spectroscopy provided the first detection of 
the Higgs mode in a BCS condensate.\cite{gia80,mas80} The definitive
identification of the absorption resonance as the $J^{\charge}=2^{+}$ Higgs mode was made by 
Avenel et al. who observed the five-fold Zeeman splitting of the zero-sound absorption resonance 
in an applied magnetic field.\cite{ave80} 

\begin{figure}[t]
\begin{center}
\includegraphics[width=\linewidth]{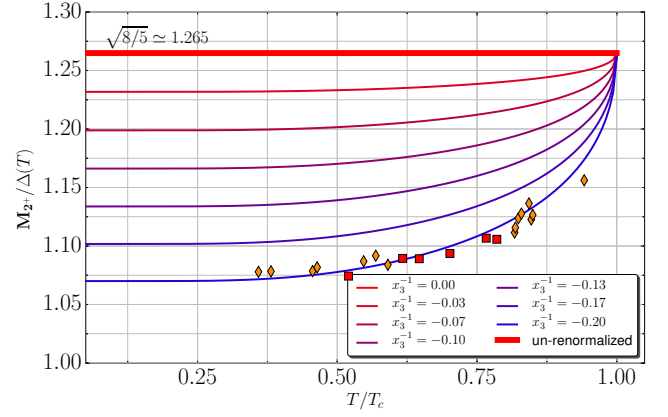}
\caption{$J^{\charge}=2^{+}$ Higgs Mass. The data are for a pressures
of $p=13\,\mbox{bar}$ (yellow diamonds),\cite{mas80} and for $p=0.8-3.5\,\mbox{bar}$ 
(red squares).\cite{ave80} Theoretical calculations of the mass are for $F_2^a=-0.88$
and values of the f-wave interaction in the Cooper channel given in the legend.
}
\label{fig-J=2+_vs_F2a-x3i}
\end{center}
\end{figure}

Acoustic spectroscopy provides precision measurements of the mass of the $J^{\charge}=2^{+}$ Higgs
mode. The magnitude of the polarization correction to the the $J^{\charge}=2^{+}$ Higgs mass 
for $T\rightarrow 0$ is measured to be $\delta M_{2^+}\approx -0.19\,\Delta$, as shown in 
Fig. \ref{fig-J=2+_vs_F2a-x3i}, indicating that the interactions giving rise to the mass shift are
net attractive. The data are from Ref. \onlinecite{mas80} for a pressure
of $p=13\,\mbox{bar}$ (yellow diamonds), and from Ref. \onlinecite{ave80} for pressures, 
$p=0.8-3.5\,\mbox{bar}$ (red squares). Also shown are theoretical results for the polarization
correction calculated as a function of temperature. In this case we assumed the most attractive
estimate for the exchange interaction, $F_2^a=-0.88$,\cite{hal90}, which accounts for
only half of the measured value of $\delta M_{2^+}$. An attractive f-wave interaction,
$x_3^{-1}\approx -0.2$, in the Cooper channel provides the additional polarization correction.
If we use the weaker value of $F_2^{a} \simeq -0.37$ reported by the 
\emph{Helium-Three Calculator}\cite{har00} for $p=13\,\mbox{bar}$ we obtain a correspondingly
stronger attractive f-wave interaction, $x_3^{-1}\simeq -0.35$.
An attractive f-wave interaction of similar magnitude, $x_3^{-1}\simeq -0.33$ at $p\approx 4.3\,\mbox{bar}$, 
is also inferred from an analysis of acoustic Faraday rotation of transverse sound that is mediated by the 
$J^{\charge}=2^{-}$ Higgs mode.\cite{lee99,sau00a} Analysis of recent acoustic Faraday rotation 
measurements, outside the regime of the linear Zeeman splitting of the 
energy levels of the $J^{\charge}=2^{+}, m$ modes, report comparable or smaller values: 
$x_3^{-1}\approx -0.4$ to $x_3^{-1}\approx -0.2$.\cite{col13,dav06}
A complete and systematic determination of the relevant interactions in the Landau and Cooper channels 
is possible from the combined measurements of the masses of the $J^{\charge}=2^{\pm}$ modes 
using longitudinal and transverse sound spectroscopy, combined with measurements of the velocities of 
zero-sound, first-sound, and the magnetic susceptibilties in both the normal- and superfluid phases of \He.

%-----------------------------
%-----------------------------

\vspace*{-3mm}
\subsection{Light Higgs Modes in the $J^{\charge}=1^+$ Sector}\label{sec-Light_Higgs}
\vspace*{-3mm}

The $J^{\charge}=1^{+}$ mode amplitudes can be related to the parameters of the degeneracy space
of {\it relative} spin- and orbital rotations, i.e. $\cR[\vartheta\vn]\in\point{SO(3)}{L-S}$, 
where $\vn$ is the axis of rotation, defined by polar and azimuthal angles, and a third variable being 
angle of rotation, $\vartheta$.
The angles define massless NG modes reflecting the spontaneous breaking of {\it separate} 
symmetries under spin- and orbital rotations, i.e. $\point{SO(3)}{L}\times\point{SO(3)}{S}$.
The $J^{\charge}=1^{+}$ multiplet provides a novel example of mass generation corresponding to the 
``Light Higgs'' extension of the standard model in particle physics.\cite{zav16} 
The Light Higgs scenario works as follows:
In \He\ separate invariance under spin- and orbital rotations is broken by the nuclear dipole-dipole 
interaction, which acts as weak symmetry breaking perturbation with an energy scale of order 
$V_{\text{D}}\sim 10^{-7}\,\mbox{K}$ per particle compared to the characteristic two-body interaction 
energy of order $V\sim 1\,\mbox{K}$.
The dipolar energy lifts the degeneracy with respect to separate spin and orbital rotations, which renders 
the $J^{\charge}=1^{+}$ multiplet a triplet of ``pseudo Nambu-Goldstone modes'' in which one or more of the 
NG modes acquires a mass from the weak symmetry breaking field. 
Long wavelength excitiations of the axis of rotation, $\vn$, remain gapless; however, excitations of the 
rotation angle, $\vartheta$, acquire a mass gap 
$M_{\text{LH}}/\hbar$ $=$ $\Omega_{\text{B}}\simeq 10\,\mbox{kHz}\ll 2\Delta/\hbar\simeq 100\,\mbox{MHz}$, 
where $\Omega_{\text{B}}$ is the longitudinal NMR resonance frequency of \Heb.
An external magnetic field further lifts the degeneracy of the remaining zero mass NG modes which split into an
optical magnon with mass, $M_{\text{opt}} = \hbar\gamma B$, and a massless acoustic magnon.
A direct detection of the Light Higgs Boson in \Heb\ was recently achieved by measuring the decay of optical 
magnons created by magnetic pumping (a magnon BEC). A sharp threshold for decay of optical magnons
to a pair of Light Higgs modes was observed by tuning the mass of the optical magnons on resonance, i.e. 
$M_{\text{opt}} = \hbar\gamma B\ge 2M_{\text{LH}}=2\hbar\Omega_{\text{B}}$.\cite{zav16}

\vspace*{-5mm}
\section{Summary and Outlook}
\vspace*{-3mm}

Mass generation based on spontaneous symmetry breaking and the introduction of an internal 
symmetry (particle-hole symmetry in BCS theory) implies a connection between the masses of the 
Fermion and Boson excitations of the broken symmetry vacuum state, and a hidden supersymmetry
in the class of BCS-NJL theories.\cite{nam85,nam95}
Nambu's proposed sum rule, inspired in part by the Bosonic spectrum of \Heb,  however, is not protected 
against symmetry breaking perturbations to the broken symmetry vacuum state, including polarization of 
the vacuum state by excitation of a Higgs boson with symmetry distinct from that of the vacuum.
For the case of \Heb, we show that corrections to the weak-coupling BCS theory and Fermionic 
interactions combined with vacuum polarization by the Higgs fields, lead to corrections
to the masses of the Higgs modes, and in general a violation of the NSR.
Our results, as well as other effects of weak perturbations like the nuclear dipolar energy, the 
Zeeman energy and weak violations of particle-hole symmetry, highlight the roles of symmetry 
breaking perturbations.

Current research in topological condensed matter addresses the transport properties and
spectrum of Fermionic excitations confined near surfaces, interfaces and edges of topological 
insulators and topological superconductors. Relatively recent theoretical work has shown how 
supersymmetry can also emerge at the boundary of topological superfluids.\cite{gro14} 
The B-phase of superfluid \He\ is the realization of a 3D time-reversal invariant topological 
superfluid, with a spectrum of helical Majorana Fermions confined on any bounding surface.
Thus, a frontier in topological quantum fluids is the role of 
confinement as a symmetry breaking perturbation on the Bosonic spectrum of confined \Heb, 
and the possible signatures of the surface spectrum of Majorana Fermions in the Bosonic modes of 
confined \Heb. New studies of the effects of confinement and symmetry-breaking perturbations on both 
the bulk and surface Bosonic and Fermionic excitations of topological superfluids will hopefully 
shed new light on the connection between spontaneous symmetry breaking, hidden supersymmetry and 
topology of the broken symmetry vacuum state in topological superfluids.

\vspace*{-3mm}
\section{Acknowledgements}
\vspace*{-3mm}

The research of JAS was supported by the National Science Foundation (Grants DMR-1106315 and DMR-1508730). 
The work of JAS was also carried out in part at the Aspen Center for Physics with partial support by 
National Science Foundation grant PHY-1066293.
The work of T. M. was supported by JSPS (No.~JP16K05448) and ``Topological Materials Science'' (No.~JP15H05855) 
KAKENHI on innovation areas from MEXT.
We thank Chandra Varma, Grigory Volovik and Anton Vorontsov for discussions on the spectrum of collective 
modes in superfluid \He{} and unconventional superconductors that informed this work.

%\vspace*{-3mm}
\section{Appendix}\label{sec-appendix}
%\vspace*{-3mm}

\subsection{TDGL Effective Potentials}\label{sec-appendix-TDGL_potentials}
\vspace*{-3mm}
The potentials that enter the TDGL functional that determine the masses of the 
Bosonic modes are given by $u_{p}=\Delta^2\,\bar{u}_{p}$
\ber
\label{eq-TDGL_potential-1}
\bar{u}_{1} 
&=& 
\nicefrac{4}{3}\Tr{\cD}\Tr{\cD^{*}}  + \Tr{\cD\cD^{\text{tr}}} + \Tr{\cD\cD^{\text{tr}}}^{*} 
\,,\qquad
\\
\label{eq-TDGL_potential-2}
\bar{u}_{2} 
&=& 
2\,\Tr{\cD\cD^{\dag}} + \nicefrac{1}{3}\left(\Tr{\cD} + \Tr{\cD^{*}}\right)^2
\,,
\\
\label{eq-TDGL_potential-3}
\bar{u}_{3} 
&=& 
\nicefrac{1}{3}\left(\Tr{\cD\cD^{\text{tr}}} + \Tr{\cD\cD^{\text{tr}}}^{*}  \right)
\nonumber
\\
&+&
\nicefrac{2}{3}
\left(\Tr{\cD\cD^{\dag}}
 + 
\Tr{\cD\cD^{*}}\right)
\,.
\label{eq-TDGL_potential-4}
\eer
\ber
\bar{u}_{4} 
&=& 
\nicefrac{4}{3}\Tr{\cD\cD^{\dag}}
+
\nicefrac{1}{3}\left(\Tr{\cD^2} + \Tr{\cD^2}^{*} \right)
\,,\qquad\qquad
\\
\label{eq-TDGL_potential-5}
\bar{u}_{5} 
&=& 
\nicefrac{1}{3}\left(\Tr{\cD\cD^{\dag}} + \Tr{\cD\cD^{\dag}}^{*} \right)
\nonumber\\
&+& 
\nicefrac{1}{3}\left(\Tr{\cD^{\text{tr}}\cD} + \Tr{\cD^{\text{tr}}\cD}^{*} \right)
+
\nicefrac{2}{3} \Tr{\cD\cD^{*}}
\,.
\eer
Note that these potentials are defined relative to the BW ground state, and thus invariant only under 
$\point{SO(3)}{J}\times\time$.

\vspace*{-3mm}
\subsection{Symmetry Relations}\label{sec-appendix-symmetries}
\vspace*{-3mm}
The components of the $4\times 4$ Nambu propagator are related by fundamental symmetries with respect
to (i) permutation exchange symmetry and (ii) conjugation symmetry. These symmetries imply the
following relations between the components of the quasiclassical propagator, 

%\eject
%\vfill

\vspace*{-3mm}
\subsubsection{Exchange symmetry}\label{sec-appendix-propagator_exchange-symmetry}
\vspace*{-3mm}
\ber
g'(\hat{p},\varepsilon_n;\vq,\omega_m) &=& +g(-\hat{p},-\varepsilon_n;\vq,\omega_m)
\,,
\\
\vec{g}'(\hat{p},\varepsilon_n;\vq,\omega_m) &=& +\vec{g}(-\hat{p},-\varepsilon_n;\vq,\omega_m)
\,,
\\
f(\hat{p},\varepsilon_n;\vq,\omega_m) &=& +f(-\hat{p},-\varepsilon_n;\vq,\omega_m)
\,,
\\
\vec{f}(\hat{p},\varepsilon_n;\vq,\omega_m) &=& -\vec{f}(-\hat{p},-\varepsilon_n;\vq,\omega_m)
\,,
\eer 
as well as for the mean-field self energies,
\ber
\Sigma'(\hat{p};\vq,\omega_m) &=& +\Sigma(-\hat{p};\vq,\omega_m)
\,,
\label{eq-exchange_symmetry-scalar}
\\
\vec{\Sigma}'(\hat{p};\vq,\omega_m) &=& +\vec{\Sigma}(-\hat{p};\vq,\omega_m)
\,,
\label{eq-exchange_symmetry-vector}
\\
d(\hat{p};\vq,\omega_m) &=& +d(-\hat{p};\vq,\omega_m)
\,,
\label{eq-exchange_symmetry-singlet}
\\
\vec{d}(\hat{p};\vq,\omega_m) &=& -\vec{d}(-\hat{p};\vq,\omega_m)
\,.
\label{eq-exchange_symmetry-triplet}
\eer 
Note that Eqs. \ref{eq-exchange_symmetry-singlet} and \ref{eq-exchange_symmetry-triplet} 
reflect the fact that spin-singlet Cooper pairs
have even parity, while spin-triplet pairs are odd-parity. 

\vspace*{-3mm}
\subsubsection{Conjugation symmetry}\label{sec-appendix-propagator_conjugation-symmetry}
\vspace*{-3mm}
The conjugation symmetry relations follow from complex conjugation of the two-point functions.
\ber
g'(\hat{p},\varepsilon_n;\vq,\omega_m)&=&+g(-\hat{p},\varepsilon_n;-\vq,\omega_m)^*
\,,
\\
\vec{g}'(\hat{p},\varepsilon_n;\vq,\omega_m)&=&+\vec{g}(-\hat{p},\varepsilon_n;-\vq,\omega_m)^*
\,,
\\
f'(\hat{p},\varepsilon_n;\vq,\omega_m) &=&+f(-\hat{p},\varepsilon_n;-\vq,\omega_m)^*
\,,
\\
\vec{f}'(\hat{p},\varepsilon_n;\vq,\omega_m)&=&-\vec{f}(-\hat{p},\varepsilon_n;-\vq,\omega_m)^*
\,.
\eer 
\ber
\Sigma'(\hat{p};\vq,\omega_m)&=&+\Sigma(-\hat{p};-\vq,\omega_m)^*
\,,
\label{eq-conjugation_symmetry-scalar}
\\
\vec{\Sigma}'(\hat{p};\vq,\omega_m)&=&+\vec{\Sigma}(-\hat{p};-\vq,\omega_m)^*
\,,
\label{eq-conjugation_symmetry-vector}
\\
d'(\hat{p};\vq,\omega_m) &=&+d(-\hat{p};-\vq,\omega_m)^*
\,,
\label{eq-conjugation_symmetry-singlet}
\\
\vec{d}'(\hat{p};\vq,\omega_m)&=&-\vec{d}(-\hat{p};-\vq,\omega_m)^*
\,.
\label{eq-conjugation_symmetry-triplet}
\eer

\begin{widetext}

\subsection{Dynamical Equations}\label{sec-appendix_Dynamical_Equations}

%
% Spin-Triplet Components of the Dynamical Pairing Self-Energy for Real, Unitary Spin-Triplet States
%
\ber
\vec{d}^{(-)}(\hp;\vq,\omega) &=& \int\dangle{p'}\,V^{(1)}(\hp,\hp')\,
			\Big\{
			\left[\onehalf\gamma + \onefourth(\omega^2-\eta'^2 -4|\vec\Delta(\hp')|^2)\bar\lambda(\hp')\right]\,\vec{d}^{(-)}(\hp') 
			 +
			\bar\lambda(\hp')\vec\Delta(\hp')\,(\vec\Delta(\hp')\cdot\vec{d}^{(-)}(\hp'))
\nonumber\\
			&&\hspace*{30mm}
			-\onehalf\eta'\,\bar\lambda(\hp')\,\vec\Delta(\hp')\,\Sigma^{(-)}(\hp')
			 - \onehalf\omega\,\bar\lambda(\hp')\,\vec\Delta(\hp')\,\Sigma^{(+)}(\hp')
			\Big\}
\,,
\label{eq-triplet-odd_parity}
\\
\vec{d}^{(+)}(\hp;\vq,\omega) &=& \int\dangle{p'}\,V^{(1)}(\hp,\hp')\,
			\Big\{
			\left[\onehalf\gamma + \onefourth(\omega^2-\eta'^2)\bar\lambda(\hp')\right]\,\vec{d}^{(+)}(\hp') 
			 -
			\bar\lambda(\hp')\vec\Delta(\hp')(\vec\Delta(\hp')\cdot\vec{d}^{(+)}(\hp'))
\nonumber\\
			&&\hspace*{30mm}
			 +\nicefrac{i}{2} \eta'\,\bar\lambda(\hp')\,\vec\Delta(\hp')\times\vec\Sigma^{(-)}(\hp')
			 + \nicefrac{i}{2}\omega\,\bar\lambda(\hp')\,\vec\Delta(\hp')\times\vec\Sigma^{(+)}(\hp')
			\Big\}
\,,
\label{eq-triplet-even_parity}
\eer
%\end{widetext}
%
% Scalar Components of the Dynamical Self-Energy for Real, Unitary Spin-Triplet States
%
\ber
\Sigma^{(+)}(\hp;\vq,\omega) &=& \Sigma^{(+)}_{\text{ext}}(\hp) + \int\dangle{p'}\,A^{s}(\hp,\hp')\,
			\Big[
			\left(\frac{\omega^2}{\omega^2-\eta'^2}\right)\,\left(1-\lambda(\hp')\right)\,\Sigma^{(+)}(\hp')
			+ 
			\left(\frac{\omega\eta'}{\omega^2-\eta'^2}\right)\,\left(1-\lambda(\hp')\right)\Sigma^{(-)}(\hp')
\nonumber\\
			&&\hspace*{40mm}
			+
			\onehalf\omega\,\bar\lambda(\hp')\,\vec\Delta(\hp')\cdot\vec{d}^{(-)}(\hp')
			\Big]
\label{eq-Sigma+Scalar}
\\
\Sigma^{(-)}(\hp;\vq,\omega) &=& \Sigma^{(-)}_{\text{ext}}(\hp) + \int\dangle{p'}\,A^{s}(\hp,\hp')\,
			\Big[
			\left(\frac{\omega\eta'}{\omega^2-\eta'^2}\right)\,\left(1-\lambda(\hp')\right)\Sigma^{(+)}(\hp')
			+
			\Big\{1 + \left(\frac{\eta'^2}{\omega^2-\eta'^2}\right)\,\left(1-\lambda(\hp')\right)\Big\}\,\Sigma^{(-)}(\hp')
\nonumber\\
			&&\hspace*{40mm}
			+
			\onehalf\eta'\,\bar\lambda(\hp')\,\vec\Delta(\hp')\cdot\vec{d}^{(-)}(\hp')
			\Big]
\label{eq-Sigma-Scalar}
\eer
%
% Vector Components of the Dynamical Self-Energy for Real, Unitary Spin-Triplet States
%
\ber\label{eq_Sigma-Vector-Plus}
\vec\Sigma^{(+)}(\hp;\vq,\omega) &=& \vec\Sigma^{(+)}_{\text{ext}}(\hp) + \int\dangle{p'}\,A^{a}(\hp,\hp')\,
			\Big[
			\left(\frac{\omega^2}{\omega^2-\eta'^2}\right)\,\left(1-\lambda(\hp')\right) \,\vec\Sigma^{(+)}(\hp')
			+
			\bar\lambda(\hp')\left(\vec\Delta(\hp')\cdot\vec{\Sigma}^{(+)}(\hp')\right)\vec\Delta(\hp')
\nonumber\\
			&& \hspace*{40mm}
			+
			\left(\frac{\omega\eta'}{\omega^2-\eta'^2}\right)\,\left(1-\lambda(\hp')\right)\vec\Sigma^{(-)}(\hp')
			\quad 
			- 
			\nicefrac{i}{2}\omega\,\bar\lambda(\hp')\,\vec\Delta(\hp')\times\vec{d}^{(+)}(\hp')
			\Big]
\\
\vec\Sigma^{(-)}(\hp;\vq,\omega) &=& \vec\Sigma^{(-)}_{\text{ext}}(\hp) + \int\dangle{p'}\,A^{a}(\hp,\hp')\,
			\Big[
			\Big\{1+\left(\frac{\eta'^2}{\omega^2-\eta'^2}\right)\,\left(1-\lambda(\hp')\right)\Big\}\,\vec\Sigma^{(-)}(\hp')
			-
			\bar\lambda(\hp')\left(\vec\Delta(\hp')\cdot\vec{\Sigma}^{(-)}(\hp')\right)\vec\Delta(\hp')
\nonumber\\
			&& \hspace*{40mm}
			+
			\left(\frac{\omega\eta'}{\omega^2-\eta'^2}\right)\,\left(1-\lambda(\hp')\right)\vec\Sigma^{(+)}(\hp')
			\quad
			-
			\nicefrac{i}{2}\eta'\,\bar\lambda(\hp')\,\vec\Delta(\hp')\times\vec{d}^{(+)}(\hp')
			\Big]
\,,
\label{eq_Sigma-Vector-Minus}
\eer
\end{widetext}
where $\eta'\equiv \vv_{\hp'}\cdot\vq$ and the Tsuneto function, 
$\lambda(\hp')\equiv\lambda(\eta',\omega;|\Delta(\hp')|)$, for $\vq\ne 0$ is given by Eq. (62) 
of Ref. \onlinecite{moo93}. 
The particle-particle interaction vertex in the spin-triplet channel is parametrized by an 
interaction parameter, $v_{\ell}$, for each odd-parity angular momentum channel, as in 
Eq. \ref{eq-pairing_interactions}. 
In the case of the particle-hole interaction vertex, the functions $A^{s,a}(\hp,\hp')$ are the 
forward scattering amplitudes for spin-independent ($A^{s}$) and spin-exchange ($A^{a}$) scattering of 
quasiparticles with momenta near the Fermi surface. These amplitudes are related to the 
the Landau interactions, $F^{s,a}(\hp,\hp')$, by the integral equation,

\be\label{eq_forward-scattering_Landau}
A^{s,a}(\hp,\hp') = 
F^{s,a}(\hp,\hp') - 
\int\frac{d\Omega_{p''}}{4\pi}
F^{s,a}(\hp,\hp'')\,A^{s,a}(\hp'',\hp')
\,.
\ee
The standard parametrization of the Landau interaction function in terms of the Landau parameters is 
$F^{s,a}(\hp,\hp') = \sum_{\ell\ge 0}\,F^{s,a}_{\ell}\,P_{\ell}(\hp\cdot\hp')$.

%----------------------------------------------------------------------------------------------
\newpage
%\bibliographystyle{unsrtnat-edit.bst}
%\bibliographystyle{apsrev4-1_PRX_style.bst}
%\bibliography{QFS,CM,Books,QM}
%\end{document}
\end{document}